\documentclass[lettersize,journal]{IEEEtran}
\IEEEoverridecommandlockouts
\usepackage{cite}
\usepackage{amsmath,amssymb,amsfonts}
\usepackage{graphicx}
\usepackage{textcomp}
\usepackage{xcolor}
\usepackage{cleveref}
\usepackage[caption=false,font=normalsize,labelfont=sf,textfont=sf]{subfig}
\def\BibTeX{{\rm B\kern-.05em{\sc i\kern-.025em b}\kern-.08em
		T\kern-.1667em\lower.7ex\hbox{E}\kern-.125emX}}
\usepackage{multirow}
\usepackage{booktabs}
\usepackage{amsfonts}
\usepackage{amsmath}
\usepackage{amsmath,cases}
\usepackage{amssymb}
\usepackage{blindtext, graphicx}
\usepackage{cite}
\usepackage{epsfig}
\usepackage{url}
\usepackage{algorithm}
\usepackage{algorithmicx, algpseudocode}
\usepackage{epstopdf}
\usepackage{color}
\usepackage{mathrsfs}
\usepackage[justification=centering]{caption}
\usepackage{float}
\usepackage{cuted}
\usepackage{lipsum}
\usepackage{diagbox}
\usepackage{graphicx,tabularx}
\usepackage{lipsum}
\usepackage{enumitem}
\usepackage{academicons}
\usepackage{xcolor}

\newcommand{\orcid}[1]{\href{https://orcid.org/#1}{\textcolor[HTML]{A6CE39}{\aiOrcid}}}

\newtheorem{theorem}{\pmb{Theorem}}

\newcommand{\pmw}{\pmb{w}}

\newcommand{\pmW}{\pmb{W}}
\newcommand{\sigk}{\sigma_{k}}
\newcommand{\sige}{\sigma_{e}}
\newcommand{\pmt}{\pmb{\theta}}

\newcommand{\btheta}{\pmb{\theta}}
\newcommand{\thetak}{\pmb{\theta}^{(n)}}
\newcommand{\thetako}{\pmb{\theta}^{(n+1)}}

\newtheorem{mylem}{\textbf{Lemma}}

\newcommand{\la}{\langle}
\newcommand{\ra}{\rangle}

\newcommand{\Prf}{{\it Proof: \ }}

\newcommand{\clC}{{\cal C}}

\newcommand{\clR}{{\pmb{\cal R}}}

\newcommand{\wko}{\pmb{w}^{(n+1)}}

\newcommand{\wk}{\pmb{w}^{(n)}}

\newcommand{\bbC}{\mathbb{C}}

\newcommand{\upsilonk}{\upsilon^{(n)}}

\allowdisplaybreaks
\begin{document}
	\title{"Security for Everyone"  in Finite Blocklength IRS-aided Systems With Perfect and Imperfect CSI}
	\author{Monir Abughalwa 
		\IEEEmembership{Student Member, IEEE}, Diep N. Nguyen, \IEEEmembership{Senior Member, IEEE},\\ Dinh Thai Hoang, \IEEEmembership{Senior Member, IEEE}, Van-Dinh Nguyen, \IEEEmembership{Senior Member, IEEE}, \\ Ming Zeng, \IEEEmembership{Senior Member, IEEE}, Quoc-Viet Pham, \IEEEmembership{Senior Member, IEEE}, \\ and Eryk Dutkiewicz, \IEEEmembership{Senior Member, IEEE}.
		\thanks{Monir Abughalwa, Diep N. Nguyen, Dinh Thai Hoang, and Eryk Dutkiewicz are with School of Electrical and Data Engineering, University of Technology Sydney, Sydney, Australia (e-mail: monir.abughalwa@student.uts.edu.au; diep.nguyen@uts.edu.au; hoang.dinh@uts.edu.au; eryk.dutkiewicz@uts.edu.au).}
		\thanks{Ming Zeng is with Department of Electrical Engineering and Computer Engineering, Universite Laval, Quebec, QC G1V 0A6, Canada (e-mail: ming.zeng@gel.ulaval.ca).}
		\thanks{Quoc-Viet Pham is with Department of School of Computer Science and Statistics, Trinity College Dublin, Dublin 2, D02 PN40, Ireland (e-mail: viet.pham@tcd.ie).}
		\thanks{Van-Dinh Nguyen is with College of Engineering and Computer Science, VinUniversity, Vinhomes Ocean Park, Hanoi, Vietnam (e-mail: dinh.nv2@vinuni.edu.vn).}
	}	
	\maketitle
	\begin{abstract}
			Provisioning secrecy for \emph{all users}, given the heterogeneity in their channel conditions, locations, and the unknown location of the attacker/eavesdropper, is challenging and not always feasible. The problem is even more difficult under finite blocklength constraints that are popular in ultra-reliable low-latency communication (URLLC). This work takes the first step to guarantee secrecy for all URLLC users in the finite blocklength regime (FBR) where intelligent reflecting surfaces (IRS) are used to enhance legitimate users' reception and thwart the potential eavesdropper (Eve) from intercepting. To this end, we aim to maximize the minimum secrecy rate (SR) across all users by jointly optimizing the transmitter’s beamforming vectors and the IRS’s passive reflective elements (PREs) under FBR latency constraints. In the FBR-SR expression, the square-root dispersion factor under the FBR, couples with the latency and reliability constraints, making the problem significantly more challenging than the same problem in the long blocklength regime (LBR). Under imperfect CSI, the problem becomes even more challenging due to nonconvex, semi-infinite constraints arising from channel estimation errors. To address this, we approximately linearize the objective function, constraints and decompose the problem into sequential subproblems. We then apply the successive convex approximation (SCA), and the $\mathcal{S}$-procedure approach to transform imperfect CSI-related semi-infinite constraints into finite linear matrix inequalities (LMIs) that can be efficiently solved. We prove that our proposed algorithms converge to a locally optimal solution with low computational complexity thanks to our closed-form linearization approach. This makes the solution scalable for large IRS deployments.  Extensive simulations with practical settings show that our approach can ensure secure communication for all users while satisfying FBR constraints even with only imperfect CSI.
		\end{abstract}
		\section{Introduction} 
		Intelligent reflective surface (IRS) has emerged as a promising technology, garnering significant interest due to its ability to enhance/tailor the passive radio environment. IRSs consist of passive reflective elements (PREs) equipped with phase shift controllers, allowing them to intelligently manipulate reflected signals \cite{nguyen2022leveraging}. By optimizing signal reflections, IRSs improve reception by creating favorably reflected multi-path signals at the receivers. Thanks to their cost-effective design and convenient deployment typically on the facades of high-rise buildings, IRSs hold immense potential for various applications, particularly in urban areas where line-of-sight channels between transmitters (Tx) and receivers (Rx) frequently face obstructions \cite{abughalwa2022finite}. Particularly, for the Internet of Things (IoT) devices that have limited computing capability and battery/energy, the IRS has gained paramount attention aiming to enhance both spectral and energy efficiency \cite{9896755}.
		\subsection{Related Works and Motivations}
		Another potential application of IRSs is to enhance the security/privacy of users by purposely manipulating reflected signals from the Tx to facilitate the signal reception at legitimate users while maximizing the multi-user interference/degrading the signals at potential eavesdroppers. In \cite{zhou2021secure}, the authors studied the design of a single user IRS-aided system to maximize the legitimate user's secrecy rate (SR), which is defined as the difference between the rate of the legitimate channel and that of the channel from the transmitter to a potential eavesdropper. The authors maximize the single-user SR by jointly optimizing the beamformer at the transmitter and the IRS. 
        When the channel state information (CSI) from the IRS to the users/receivers is unknown or imperfect, the authors in \cite{10256584} studied an IRS-aided multi-user system, where they formulated a secrecy sum rate (SSR) maximization problem with the eavesdropper's channel partially known to the receiver. It is worth mentioning that the SSR problem does not guarantee the secrecy for \emph{all users}.	
		
	In parallel, low-latency communication (URLLC), which relies on the finite blocklength (short packet) regime of communication (FBR), has been envisioned as a key technology to serve the critical missions of IoT. Since FBR often requires a stricter design approach than the long blocklength regime/communication (LBR) systems\cite{durisi2016toward}, maintaining high-reliability communication is more challenging due to the lower channel coding gain. Moreover, in URLLC applications such as intelligent transportation, leakage information could expose the user's location or identity. As a potential solution to secure URLLC, the use of IRS has recently attracted paramount interest. IRS links exhibit low latency that is beneficial for the FBR strict requirements \cite{10904325}.  Zhao et al. studied the information freshness in a single-user FBR-IRS-aided system with the presence of an eavesdropper in \cite{10904325}. The authors derived a closed-form expression for the upper bound secrecy outage probability under statistical CSI. The single user's secrecy outage is minimized by jointly optimizing the blocklength and the IRS PREs. 
    The authors in \cite{10068292} studied the secrecy performance of an IRS-assisted URLLC system, where an FBR coding scheme was proposed to secure the link, showing that secrecy is achievable beyond a certain blocklength threshold. However, it does not consider the joint optimization of secrecy rate and transmission parameters, which is crucial for practical deployment. Nevertheless, the assumption that Alice/Tx can obtain Eve's CSI, as stated before, is not always feasible. Note that most earlier work focused only on single-user systems. The SSR maximization approach in LBR-IRS systems is neither applicable to nor can it guarantee the secrecy for all users in FBR-IRS systems due to the additional latency requirements of the FBR.
   
	Guaranteeing secrecy in FBR is significantly more challenging than in conventional LBR applications. Unlike LBR, where sufficiently long codewords can asymptotically drive both error probability and leakage to zero, these factors remain strictly non-zero in FBR due to the coding limits imposed by short blocklengths \cite{6802432}. In particular, the square-root of the channel dispersion in the FBR-SR expression directly captures the effect of finite blocklength on the decoding error probability, making the SR a non-linear function. Consequently, the optimization problem is significantly more complex, as these non-zero error and leakage terms introduce additional nonconvex penalties that must be jointly considered with power and beamforming constraints.  Motivated by the above, this work takes the first step to guarantee secrecy for all URLLC users in FBR-IRS-aided systems. Unlike existing work, which only considers secrecy enhancement in LBR-IRS-aided systems, or FBR secrecy improvement with full CSI \cite{10082954}, we tackle the problem of secrecy guarantee in FBR-IRS-aided systems under perfect/imperfect IRS to users'/Eve's CSI. To this end, we consider a popular case where IRSs are used to enhance/aid the signal reception at legitimate users (from a transmitter) under the presence of a potential eavesdropper, while the direct/line-of-sight channel between the transmitter to all users and eavesdropper does not exist (e.g., due to blockages).\footnote{In reality, it is impractical to consider the underlying problem if the eavesdropper has a line-of-sight channel with the transmitter while the users do not. In such cases, a friendly jammer should be used to protect users' privacy \cite{9402750}.} We then maximize the minimum SR among all the users given the FBR constraints by optimizing the transmit beamforming vector and the IRS's PREs. We consider both cases with perfect and imperfect CSI from the IRS to the users and the eavesdropper. For the perfect CSI, we assume that the transmitter can obtain the correct CSI of all users and the eavesdropper. This is the case when one of the legitimate users gets compromised and acts as an eavesdropper, trying to eavesdrop on other users; hence, its CSI would be available to Alice.

	When only partial or erroneous CSI (from the IRS to the users and Eve) is available, the problem becomes even more challenging. In particular, the semi-infinite constraints introduced by imperfect CSI are embedded within the dispersion term of the FBR-SR expression, which involves both the inverse Q-function and the square-root of the channel dispersion. This embedding makes the decoding error probability highly sensitive to channel estimation errors \cite{7529226}, while the square-root further increases the nonlinearity. To tackle the problem, we first introduce slack variables to deal with the transmit beamforming vectors and the IRS's PREs coupling within the objective function. Secondly, we leverage the successive convex approximation (SCA) technique \cite{9197675}, and the $\mathcal{S}$-procedure \cite{boyd1994linear} to convert the semi-infinite constraints into linear matrix inequalities (LMI). Although the SCA framework with the $\mathcal{S}$-procedure converts the semi-infinite constraints into LMIs, they remain nonconvex. We address this by applying a first-order Taylor approximation to linearize the nonconvex components, yielding a convex reformulation solvable at each iteration. Then, we propose a penalty convex-concave procedure (PCCP) \cite{9525400} to tackle the unit modulus constraint (UMC) of the IRS. Extensive simulations with practical settings show that maximizing the minimum SR among all the users can achieve a better chance of ensuring secure communications for all users (subject to the location of the eavesdropper) under the FBR constraints even under imperfect CSI conditions. 
	\subsection{Contributions}
	The main contributions are summarized as follows: 
	\begin{itemize}
		\item We take the first step to guarantee secrecy for all URLLC users in the finite blocklength regime (FBR) where IRSs are used to enhance legitimate users' reception while thwarting the potential Eve from intercepting. To that end, we aim to maximize the minimum SR among all users by jointly optimizing the transmitter’s beamforming and the IRS’s passive reflective elements (PREs), while meeting the FBR latency constraints. The resulting optimization problem is nonconvex and even more complicated under imperfect channel state information (CSI). 
		\item When the CSI is available, we tackle the above nonconvex problem by linearizing its objective function with tractable approximation functions, leading to a computationally efficient algorithm. To tackle UMC, unlike traditional methods, i.e., conventional SDR, we directly optimize the PREs arguments to provide a low-complexity solution that can scale with large IRSs. The resulting solution is proved to converge to a locally optimal solution of the original nonconvex problem.
		\item When perfect CSI is not available, we use the SCA approach to transform imperfect CSI-related semi-infinite constraints into finite LMI. We prove that our proposed algorithm converges to a locally optimal solution with low computational complexity thanks to our closed-form linearization approach. This makes the solution scalable for large IRS deployments.
		\item Eventually, we perform extensive simulations with practical settings. The simulation results show that our approach can ensure secure communication for all users while satisfying FBR constraints even under imperfect CSI.  Additionally, the impact of the reflected channel's imperfect CSI on the system is also evaluated.
	\end{itemize}
	The remainder of this paper is organized as follows. The system model is discussed in Sections \ref{sec2}. Then, Sections \ref{sec3} and \ref{sec4} present the problem and corresponding solutions under perfect and imperfect CSI, respectively. The extensive simulations and discussion are in Section \ref{sec5}. Finally, Section \ref{conc} concludes the paper.
	
	This paper uses the following notation: bold letters denote vectors and matrices. $\pmb{I}_M$ denotes an M dimensional identity matrix. Diag($m_1,\dots,m_m$) denotes the diagonal matrix with diagonal entries of $\{m_1,\dots,m_m\}$. The symbols $\Re$ and $\mathbb{C}$ represent the real and the complex field, respectively. $\clC(0,\bar{z})$ denotes the circular Gaussian random variable with zero mean and variance $\bar{z}$. For matrices $\mathbf{C}$ and $\mathbf{D}$, $\la \mathbf{C},\mathbf{D} \ra \triangleq \mbox{trace}\left(\mathbf{C}^H \mathbf{D}\right)$. For matrix $\mathbf{A}$, $\Re\left\{\mathbf{A}\right\}$ denotes the real part, $\la \mathbf{A}\ra \triangleq\mbox{trace}(\mathbf{A})$, the symbol $\|\mathbf{A}\|_1$ denotes the 1-norm, $\|\mathbf{A}\|$ denotes the Frobenius norm, $\mathbf{A}^*$ denotes the conjugate, $\mathbf{A}^H$ denotes the Hermitian (conjugate transpose), $\lambda_{\text{max}}\left(\mathbf{A}\right)$ denotes the maximal eigenvalue, $\angle(\mathbf{A})$ denotes its argument, and $\mathbf{A} \succeq 0$ means positive semi-definite.  
	\section{{System Model}} \label{sec2}
	\begin{figure}[!htb]
		\centering
		\includegraphics[width=0.36\textwidth]{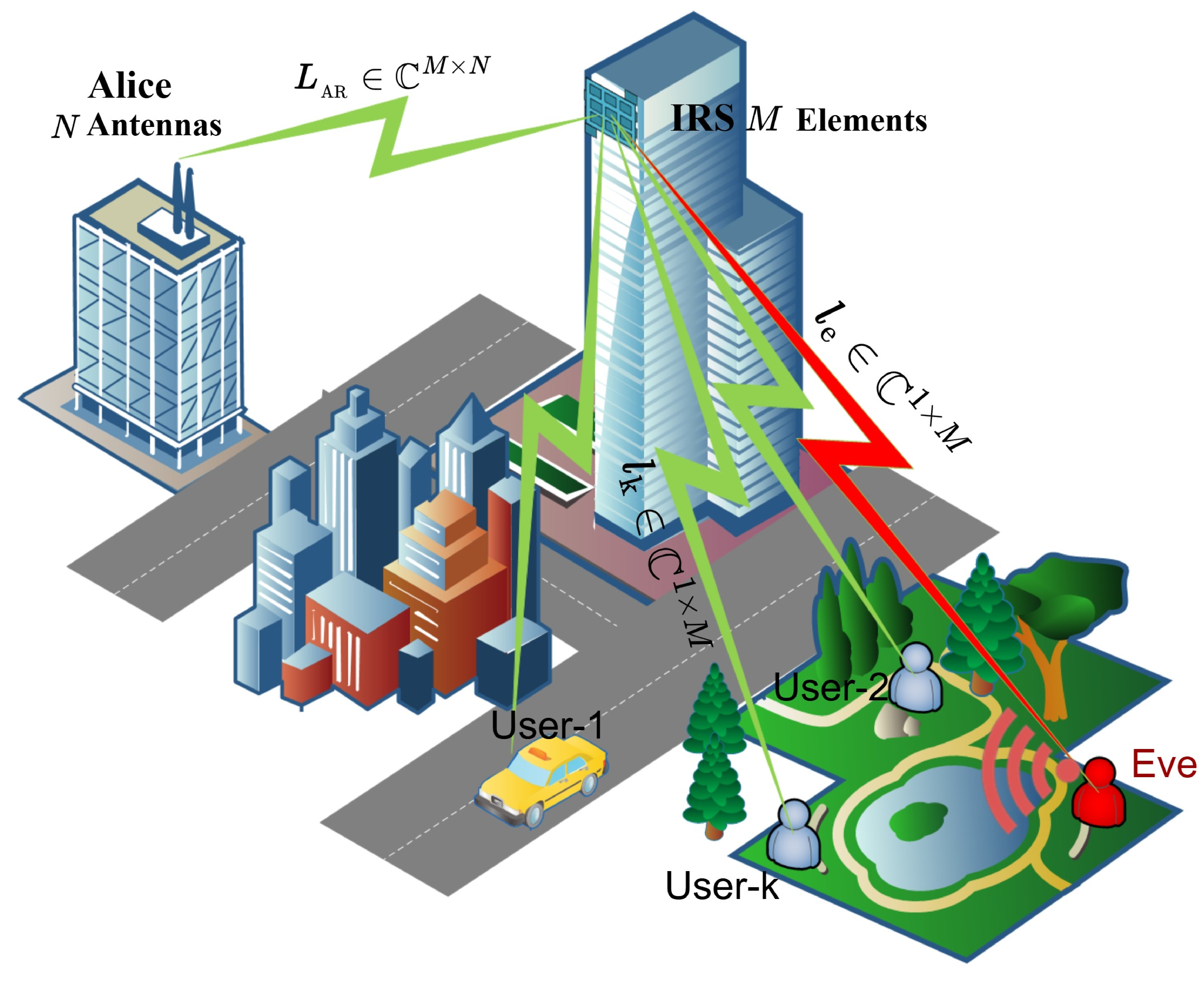}
		\caption{System model.}
		\label{s1}
	\end{figure}
	We consider an FBR-IRS-aided downlink system as depicted in Fig. \ref{s1}, in which an $N$-antenna BS (Alice) transmits confidential information to $\mathcal{K}$ single antenna users under the presence of a single antenna eavesdropper (Eve). Here, we assume the direct radio links between Alice and the users are severely blocked{\footnote{The effectiveness of the IRS in enhancing users' rate/confidentiality is limited when a strong direct link between the transmitter and the receiver exists \cite{9963699}.}}. This scenario is usually the case in highly populated areas with high-rise buildings. An IRS with $M$ PREs is thus deployed (e.g., at the facade of a building) to support the transmission between Alice and the users. Let $k \triangleq \{1,2, \dots, \mathcal{K}\}$ denote the set of the legitimate users, and $e$ denote the Eve in the system. Let 
	$\pmb{L}_{\text{AR}}= \sqrt{\aleph_{\text{{AR}}}} \tilde{{\pmb{L}}}_{_{\text{{AR}}}}\in \mathbb{C}^{M \times N}$ denote the channel from Alice to the IRS, while $\tilde{{\pmb{L}}}_{_{\text{{AR}}}}$ is modeled by Rician fading, and $\sqrt{\aleph_{\text{{AR}}}}$ is the large scale fading factor of the Alice-to-IRS link. The channel from the IRS to user-$k$ and to Eve is modeled by $\pmb{l}_i=  \sqrt{\aleph_{\text{R}i}} \tilde{\pmb{l}}_i \clR^{1/2}_{\text{R}i}\in \mathbb{C}^{1\times M}$, where $i \in \{k,e\}$, $\aleph_{\text{R}i}$ is the large-scale fading factor of the IRS-to-$i$ \cite{kammoun2020asymptotic}, $\tilde{\pmb{l}}_i$ is modeled by Rician fading, and $\clR_{\text{R}i} \in \bbC^{M\times M}$ is the IRS elements' spatial correlation matrix \cite{kammoun2020asymptotic}. For the confidential message intended for user-$k$, i.e., $s_k$, the signal received by user-$k$ and Eve, corresponding to the intended user, can be respectively expressed by:
	\begin{align} \label{chkE}
		g_i\triangleq \pmb{h}_{i}(\btheta) {\sum}_{k=1}^{\mathcal{K}} \pmw_k s_k +\bar{n}_i, i\in \{k,e\},
	\end{align}
	where  $\pmb{h}_{i}(\btheta) \in \mathbb{C}^{1 \times N}$ is the cascaded channel gain from Alice to $i \in \{k,e\}$, $\pmw_k \in \mathbb{C}^{N \times 1}$ is the beamforming vector applied for user-$k$,  $\btheta=(\theta_1,\dots, \theta_m)^T\in [0,2\pi)^M$ denotes the phase shift vector the IRS's PREs, $\bar{n}_i$ is the zero-mean Additive White Gaussian Noise (AWGN) with power density $\sigma_i$ for $i \in \{k,e\}$. The cascade channel gain $\pmb{h}_{i}(\btheta)$ from Alice to $i \in \{k,e\}$, can be written in terms of the channel gain from Alice to the IRS and the channel gain from the IRS to the user or Eve as follows:
	\begin{align} \label{h1} 
		\pmb{h}_{i}(\btheta)\triangleq {\pmb{l}}_i \pmb{\Phi} \pmb{L}_{\text{AR}} \triangleq {\pmb{l}}_i \sum_{m=1}^{M} \exp(j \theta_m) \pmb{\Gamma}_m \pmb{L}_{\text{AR}},
	\end{align}
	where $\pmb{\Gamma}_m$ is an $M \times M$ matrix with all zeros except for its $(n,n)$ entry which is 1, and $\pmb{\Phi} = \text{Diag}(e^{j \btheta})$.
	
	Given that the IRS is deployed at a known location, such as the facade of a high-rise building, the CSI between Alice and the IRS can be accurately estimated by exploiting the angles of arrival and departure \cite{9110587}, or through advanced channel estimation techniques \cite{9501057,9373363}. However, the reflected channel’s CSI from the IRS to the IoT users is much more challenging to obtain due to the passive nature of the IRS and the mobility/varying nature of the users’ environment and location  \cite{8579566}. For Eve, it is also hardly possible to pinpoint its location or its accurate CSI from the IRS. To account for the CSI imperfection, we adopt the bounded CSI model in which the reflected channel from the IRS to the users and Eve can be expressed as \cite{9266086}:
	\begin{subequations} \label{ICSIcha}
		\begin{alignat}{2} 
			&\pmb{l}_{i}&& \triangleq \widehat{\pmb{l}}_{i} + \Delta\pmb{l}_{i}, \forall i \in \{k,e\}, \label{ch2-h1} \\            
			&\omega_i && \triangleq \{\|\Delta\pmb{l}_{i}\|_2 \leq \Omega_{i}\}, \forall i \in \{k,e\},\label{ch2-h2}	
		\end{alignat}
	\end{subequations}
	where $\widehat{\pmb{l}}_{i}$ denotes the (imperfect) estimated channel vector, $\Delta\pmb{l}_{i}$ represents the channel estimation error of the corresponding estimation, $\omega_i$ is a set for all possible channel estimation errors, and $\Omega_{i}$ is the radii of the uncertainty regions as known to Alice. Hence, \eqref{h1} can be reformulated as:
	\begin{align} \label{chaCSIdef}
		\widehat{\pmb{h}}_{i}(\btheta)\triangleq \left(\widehat{{\pmb{l}}}_i+ \Delta\pmb{l}_{i}\right) \pmb{\Phi} \pmb{L}_{\text{AR}}.
	\end{align}
	In the sequel, we deal with perfect and imperfect CSI from the IRS to the users and Eve. For the former, we assume that the CSI of all the users and Eve can be accurately obtained by Alice \cite{niu2021weighted,chu2020secrecy}. As aforementioned, in this case we assume that Eve acts as a legitimate user in one period/slot but then tries to eavesdrop on other users.
	Using different well-established channel estimation methods, such as the anchor-assisted channel estimation approach \cite{9603291}, Alice can practically obtain the CSI of all the users. In this case, the reflected channel from the IRS to the users and Eve can be captured by setting $\Delta\pmb{l}_{i}$ to zero in \eqref{ch2-h1}, i.e., $\pmb{l}_{i}= \widehat{\pmb{l}}_{i}, i \in \{k,e\}$. In the second case, only partial/imperfect CSI $\widehat{\pmb{l}}_{i}$ is available. In this case, one can vary the magnitude of $\Delta\pmb{l}_{i}$, i.e., the radii $\Omega_{i}$ of the uncertainty region to capture different levels of CSI imperfection. 
	
	The corresponding signal-to-interference-plus-noise ratio (SINR) of the received signal at the user-$k$ and Eve under perfect and imperfect IRS to users/Eve CSI can be written as, respectively,
	\begin{align} \label{SINRex}
		\gamma_i(\pmw,\btheta)&\triangleq{|\pmb{h}_{i}(\btheta)\pmw_k|^2}/{\rho_{i}}, ~ i \in \{k,e\}, \\
		\widehat{\gamma}_i(\pmw,\pmt)&\triangleq{|\widehat{\pmb{h}}_{i}(\btheta)\pmw_k|^2}/{\widehat{\rho}_{i}}, ~ i \in \{k,e\},
	\end{align}
	where  $\rho_{i}\triangleq {\sum}_{j=1,j \neq k}^{\mathcal{K}}|\pmb{h}_{i}(\btheta)\pmw_j|^2+\sigma_i$, and $\widehat{\rho}_{i}\triangleq {\sum}_{j=1,j \neq k}^{\mathcal{K}}|\widehat{\pmb{h}}_{i}(\btheta)\pmw_j|^2+\sigma_i$.
	Under the presence of Eve, the closed-form expressions for the FBR-SR of the user-$k$ under perfect and imperfect IRS to users/Eve CSI are defined as, respectively, \cite[Eq.115]{8665906}:
	\begin{align}  
			\mathcal{S}_{k}^{\mathcal{F}}(\pmw,\pmt)&\triangleq \left[C_{k}-C_{e} -\xi_k \sqrt{V_{k}} - \xi_{e} \sqrt{V_{e}}\right]^+, \label{Cstot2} \\
			\widehat{\mathcal{S}}_{k}^{\mathcal{F}}(\pmw,\pmt)&\triangleq \left[\widehat{C}_{k}-\widehat{C}_{e}-\xi_k \sqrt{\widehat{V}_{k}} - \xi_{e} \sqrt{\widehat{V}_{e}}\right]^+, \label{Cstot2ICSI} 
	\end{align}
	where $[x]^+ \triangleq \text{max}[0,x]$, while the user's data rate, and the eavesdropping rate of Eve decoding $s_k$, are respectively given by,
	\begin{align} 
		C_{i}(\pmw,\pmt)&\triangleq\ln\left(1+\gamma_i(\pmw,\btheta)\right), i \in \{k,e\}.  \\
		\widehat{C}_{i}(\pmw,\pmt)&\triangleq\ln\left(1+\widehat{\gamma}_i(\pmw,\btheta)\right), i \in \{k,e\} \label{ckedef}, 
	\end{align} 
	$\xi_i \triangleq {Q^{-1}(\tau_i)}/{\text{ln}(2)\sqrt{N_t}}$, $Q^{-1}(\cdot)$ is the inverse of the Gaussian Q-function, $N_t \triangleq \mathcal{B} t_t$ is the packet length, $\mathcal{B}$ is the bandwidth, $t_t$ is the transmission duration, $\tau_k$ is the decoding error probability, $\tau_e$ is the information leakage\cite{feng2021reliable}, and $V_{k}$ and $V_{e}$ are the dispersion factors defined as \cite{feng2021reliable}, 
	\begin{align} 
		V_{i}(\pmw,\pmt) &\triangleq \frac{2 \gamma_i(\pmw,\pmt)}{(1+\gamma_i(\pmw,\pmt))} \triangleq 2\left(1-({\rho_{i}}/{\upsilon_{i}})\right), \label{dispfa1} \\
		\widehat{V}_{i}(\pmw,\pmt) &\triangleq \frac{2 \widehat{\gamma}_i(\pmw,\pmt)}{(1+\widehat{\gamma}_i(\pmw,\pmt))} \triangleq 2\left(1-({\widehat{\rho}_{i}}/{\widehat{\upsilon}_{i}})\right), \label{dispfa2}
	\end{align} 
	where $\upsilon_{i}\triangleq{\sum}_{j=1}^{\mathcal{K}}|\pmb{h}_{i}(\pmt)\pmw_j|^2+\sigma_i$, and $\widehat{\upsilon}_{i}\triangleq {\sum}_{j=1}^{\mathcal{K}}|\widehat{\pmb{h}}_{i}(\pmt)\pmw_j|^2+\sigma_i,~i\in \{k,e\}$.
		
	\vspace{1pt}Unlike the traditional SR definition in the LBR, the FBR imposes more constraints on the SR. Specifically, the reliable transmission in FBR requires that the decoding error probability $\tau_k$ at user-$k$ is not larger than the maximum decoding error probability in FBR  $\tau_{\max}$, the information leakage constraint imposes that the information leakage $\tau_e$ does not exceed $\tau_{e_{\max}}$, the maximum information leakage in FBR \cite{feng2021reliable}. Additionally, the transmission duration $t_t$ doesn't exceed the maximum transmission duration $t_{\max}$. Moreover, ensuring secure communication in the FBR regime with packet length $N_t$, is more challenging than in its LBR counterpart. This increased difficulty arises from the inherent limitations on the achievable SR in the FBR setting, which are due to the combined effects of non-negligible decoding error probabilities and information leakage. These two conditions distinguish the FBR case from its LBR counterpart. One may notice that when the transmission duration $t_t $ approaches infinity, the dispersion factor $V_i$ reaches zero, and the FBR definitions \eqref{Cstot2} and \eqref{Cstot2ICSI} reduce to the traditional LBR-SR case, 
	which can be expressed as \cite{bloch2008wireless}: 
	\begin{align} \label{Cstot}
		\mathcal{S}_{k}^{\mathcal{L}}(\pmw,\btheta)& \triangleq [C_k(\pmw,\pmt)-C_e(\pmw,\pmt)]^+, \\
		\widehat{\mathcal{S}}^{\mathcal{L}}_{k}(\pmw,\btheta)& \triangleq [\widehat{C}_k(\pmw,\pmt)-\widehat{C}_e(\pmw,\pmt)]^+.  
	\end{align}
	\section{MAXMIN SR under perfect CSI in FBR systems} \label{sec3}
	 In this section, we first address the problem of maximizing the minimum SR among all the users under the perfect CSI assumption to provide secure communications for all the users. The optimization problem can be formally stated as: 
\begin{subequations} \label{P1}
	\begin{alignat}{3} 
		&(\mathcal{P}1):&& ~\underset{\pmw,\btheta}{\max}  ~ \underset{k \in \mathcal{K}}{\min} ~ \mathcal{S}^{\mathcal{F}}_{k}(\pmw,\btheta), \label{P1-1} \\            
		&~\text{s.t.} && ~{\sum}_{k = 1}^{\mathcal{K}}\Vert \pmw_k \Vert^2 \leq P,\label{P1-2} \\
		& && |e^{(j \btheta)}| = 1,  \label{P1-3}\\
		& && \tau_k\leq \tau_{\max}, ~\tau_e \leq \tau_{e_{\max}}, ~t_t \leq t_{\max},  \label{P1-4}
	\end{alignat}
\end{subequations}
where $P$ is Alice's power budget, $\eqref{P1-2}$ captures the sum of the transmitted power constraint, $\eqref{P1-3}$ captures the UMC of the PREs' phase shift, and \eqref{P1-4} captures the maximum decoding error probability, the maximum information leakage, and the maximum transmission duration, respectively. It is worth noting that, the FBR constraints shown in \eqref{P1-4}, are inherently embedded within the objective function, since the SR expression $\mathcal{S}^{\mathcal{F}}_{k}(\pmw,\btheta)$ incorporates the square root of the dispersion factors, as shown in~\eqref{Cstot2}. As a result, the optimization problem falls outside the scope of standard convex formulations and is more challenging to solve than its LBR counterpart.


The optimization problem ($\mathcal{P}1$) is nonconvex since the objective function \eqref{P1-1} is not concave  the UMC \eqref{P1-3} is nonconvex. To tackle this nonconvex problem, one can employ the AO technique \cite{nguyen2022leveraging}. Specifically, at iteration $(n)$, the feasible point $(\wk,\thetak)$ is generated from ($\mathcal{P}1$) by solving two sub-problems, the first one is to optimize $\pmw$ with a fixed $\btheta$:
\begin{align}  \label{probw}
	(\mathcal{P}1.1): ~ \underset{\pmw}{\max} ~ \underset{k \in \mathcal{K}}{\min} ~ \mathcal{S}_{k}^{\mathcal{F}}(\wk,\btheta), ~ \text{s.t.} ~ (\ref{P1-2}),
\end{align} 
then, we optimize $\btheta$ with a fixed $\pmw$ by solving the following problem
\begin{align} \label{probth}
	(\mathcal{P}1.2): ~ \underset{\btheta}{\max} ~ \underset{k \in \mathcal{K}}{\min} ~ \mathcal{S}_{k}^{\mathcal{F}}(\pmw,\thetak), ~ \text{s.t.} ~ (\ref{P1-3}).
\end{align}	
However, with AO, solving two subproblems $(\mathcal{P}1.1)$ and $(\mathcal{P}1.2)$ is computationally demanding, especially given the large number of PREs of the IRS.  Our proposed linearization method in the sequel uses mathematically tractable approximation functions leading to a computationally efficient algorithm that can be used for a large number of IRS's PREs.
\begin{figure*}[!t] 
	\normalsize 
	\begin{align} 
		C_e(\pmw,\thetak) &\geq x_{1,k_{e}}^{(n)}+z_{1,k_{e}}^{(n)}+2\Re \sum\limits_{j=1, j \neq k}^{\mathcal{K}}\left\{\left\la \pmb{y}_{j,k_{e}}^{(n)},\pmw_j\right\ra\right\} -\frac{\rho_{e}^{(n)}}{1+\rho_{e}^{(n)}}\left(\sum\limits_{j=1,j \neq k}^{\mathcal{K}}|\pmb{h}_{e}(\thetak)\pmw_j|^2\right) - \frac{1}{\upsilon_{e}^{(n)}}\sum\limits_{j=1}^{\mathcal{K}}|\pmb{h}_{e}(\thetak)\pmw_j|^2, \label{CEtot1} 	\tag{20} \\
		{C_k}(\wko,\pmt) & \geq {x}_{1,k}^{(n+1)}+2\Re\left\{{\sum}_{m=1}^{M}{\pmb{y}}^{(n+1)}_{k,k}(m) e^{j\theta_m} \right\}+ \left(e^{j \pmt}\right)^H {\pmb{\varphi}}_{1,k}^{(n+1)} e^{j \pmt}, \label{LBBob2x} \tag{24} \\
		C_e(\wko,\pmt)&\geq x_{1,k_{e}}^{(n+1)}+z_{1,k_{e}}^{(n+1)}+2\mathfrak{R}\left\{{\sum}_{m=1}^{M} {\pmb{y}}^{(n+1)}_{k_{e}}(m) e^{(j \theta_m)} \right\}+\left(e^{j \btheta}\right)^H {\pmb{\varphi}} _{1,k_{e}}^{(n+1)} e^{j \btheta} + \left(e^{j \btheta}\right)^H {\pmb{\varphi}} _{k_{e},j}^{(n+1)} e^{j \btheta}. \label{LBEve2x} \tag{25} 
	\end{align} 
	\hrulefill 
\end{figure*}
\subsubsection{Sub-Problem for Optimizing the Beamforming Vectors} \label{subsec1}
We fix $\btheta$ given $\wk$ and solve the problem ($\mathcal{P}1$) to obtain $\wko$ satisfying ${\mathcal{S}}^{\mathcal{F}}_k(\wko,\thetak) \geq {\mathcal{S}}^{\mathcal{F}}_k(\wk,\thetak)$. 
We start by linearizing the objective function in (\ref{P1-1}), which consists of four parts: the user-$k$'s SR $C_k(\pmw,\pmt)$, Eve's negative eavesdropping rate $C_e(\pmw,\pmt)$, the user-$k$ dispersion factor $V_k(\pmw,\pmt)$, and Eve's dispersion factor $V_e(\pmw,\pmt)$ as shown in \eqref{Cstot2}. 

First, we adopt the idea in \cite{TTN16} to convert user-$k$'s SR into a linear form. Specifically, using the inequality (\ref{fund1}) in Appendix \ref{AppA}, let's define $\pmb{\Lambda}\triangleq \pmb{h}_{k}(\btheta^{(n)})\pmw_k$, $\pmb{\digamma}\triangleq \rho_k$, $\hat{\Lambda}\triangleq \pmb{h}_{k}(\btheta^{(n)})\pmw_k^{(n)}$ and
$\hat{\digamma} \triangleq \rho_k$, hence, the users' data rate can be written as: 
\begin{equation} \label{LBBob} 
	C_k(\pmw,\thetak)\geq x_{1,k}^{(n)}+2\Re\left\{\left\la \pmb{y}_{k,k}^{(n)},\pmw_k\right\ra\right\}-z_{1,k}^{(n)}\upsilon_{_{k}}^{(n)},
\end{equation}
where
\begin{align*}
	x_{1,k}^{(n)} &\triangleq C_k(\wk,\thetak)-\gamma_k(\wk,\thetak)-\sigk z_{1,k}^{(n)}, \\
	\pmb{y}^{(n)}_{k,k} &\triangleq {\pmb{h}_{_{k}}^H (\thetak)\pmb{h}_{_{k}}(\thetak)\wk_{k}}/{\rho_{k}^{(n)}}, \\
	z_{1,k}^{(n)} &\triangleq{1}/{\rho_{k}^{(n)}}-{1}/{\upsilon_{_{k}}^{(n)}}.
\end{align*}

Second, we linearize Eve's eavesdropping rate, which can be written as: 
\begin{align}\label{EveSRt1}
	-\ln(1+\gamma_e(\pmw,\thetak))&\triangleq \overset{a_1}{\overbrace{\ln(1+\rho_{e}^{(n)})}}- \overset{a_2}{\overbrace{\ln(1+\upsilon_e^{(n)})}}.  
\end{align}
Here, we adopt the idea in \cite{niu2022joint}, where the term ($a_1$) in \eqref{EveSRt1} can be linearized by defining $\pmb{z}\triangleq \rho_{e}$ and substituting it in the inequality \eqref{fund2} in Appendix \ref{AppA}. The term ($a_2$) in \eqref{EveSRt1} can be linearized by defining $\pmb{\Upsilon}\triangleq \upsilon_e$ and substituting it in the inequality \eqref{fund3} in Appendix \ref{AppA}. Hence, Eve's eavesdropping rate can be expressed as in \eqref{CEtot1}, where 
\begin{alignat*}{2}
	&x_{1,k_{e}}^{(n)} &&\triangleq \ln\left(\rho_{e}^{(n)}\right)-\rho_{e}^{(n)}-\ln\left(\upsilon_{e}^{(n)}\right)+1, \\
	&z_{1,k_{e}}^{(n)} &&\triangleq\left(-\dfrac{\rho_{e}^{(n)}}{1+\rho_{e}^{(n)}}-\dfrac{1}{\upsilon_{e}^{(n)}}\right)\sige, \\
	&\pmb{y}_{k_{e},j}^{(n)} &&\triangleq \pmb{h}_{e}^H(\thetak)\pmb{h}_{e}(\thetak) \pmw_j^{(n)}.
\end{alignat*}
Next, we linearize the FBR dispersion factors $V_{i}(\pmw,\pmt), i \in \{k,e\}$. Using \eqref{dispfa1}, the inequalities \eqref{xy} and \eqref{xt} in Appendix \ref{AppA}, and by defining $x\triangleq V_i(\wk,\pmt)$, $\mathbf{A} \triangleq \rho_{i}$, and $B \triangleq \tau_i$, the dispersion factors can be expressed as:  \setcounter{equation}{20}
\begin{align}  \label{urllcbob}
	\xi_{i} \sqrt{V_{i}(\pmw,\thetak)} \leq   x_{2,i}^{(n)}-2 \sum\limits_{j=1, j \neq k}^{\mathcal{K}}\Re\left\{\left\la \pmb{f}_{j,i}^{(n)},\pmw_j\right\ra\right\} -z_{2,i}^{(n)}\upsilonk, 
\end{align}
where,
\begin{align*}
	x_{2,i}^{(n)}&\triangleq \xi_{i}\left(\frac{\sqrt{V_i(\wk,\thetak)}}{2}+\frac{\left(\upsilon^{(n)}_i\right)^2+\rho^{(n)}_i \sigma_{i}-2 \upsilon^{(n)}_i \sigma_{i}}{\left(\upsilon^{(n)}_i\right)^2 \sqrt{V_i(\wk,\thetak)}}\right),  \\
	\pmb{f}_{j,i}^{(n)} &\triangleq  {\xi_i [\pmb{h}_{i}(\thetak)]^2\wk_j}/{\upsilon^{(n)}_i \sqrt{V_i(\wk,\thetak)}},  \\
	z_{2,i}^{(n)} & \triangleq \dfrac{\xi_{i} \rho^{(n)}_i}{\left(\upsilon^{(n)}_i\right)^2 \sqrt{V_i(\wk,\thetak)}}.
\end{align*}

By substitution \eqref{LBBob}, \eqref{CEtot1}, and \eqref{urllcbob} into (\ref{Cstot2}), the approximated surrogate FBR-SR function can be written as:  
\begin{align} \label{SRFBRtot}
	\widetilde{\mathcal{S}}_{k}^{\mathcal{F}}(\pmw,\thetak)&\geq {x}_{k}^{(n)}+2\Re\left\{\left\la {\pmb{y}}_{k}^{(n)},\pmw_{k}\right\ra\right\}-(\pmw_{k})^H {\pmb{\psi}}^{(n)}_{k} \pmw_{k},
\end{align}
where \begin{align*}
	{x}_{k}^{(n)} &\triangleq x_{1,k}^{(n)}+x_{1,k_{e}}^{(n)}+z_{1,k_{e}}^{(n)}+x_{2,k}^{(n)}+x_{2,k_{e}}^{(n)}, \\
	{\pmb{y}}_{k}^{(n)} &\triangleq {\sum}_{j=1}^{\mathcal{K}}\pmb{y}_{j,k}^{(n)}+{\sum}_{j=1,j \neq k}^{\mathcal{K}} \left(\pmb{y}_{k_{e},j}^{(n)}-\pmb{f}_{j,k}^{(n)}-\pmb{f}_{j,k_{e}}^{(n)}\right), \\
	\pmb{\psi}^{(n)}_{k}
	& \triangleq {\sum}_{j=1}^{\mathcal{K}} z_{j}^{(n)} \pmb{h}_{j}^H(\thetak) \pmb{h}_{j}(\thetak) + z_{k_{e}}^{(n)} \pmb{h}_{e}^H(\thetak) \pmb{h}_{e}(\thetak), \\
	z_{k}^{(n)} &\triangleq z_{1,k}^{(n)}+z_{2,k}^{(n)}, \\
	z_{k_{e}}^{(n)} &\triangleq {\rho_{e}^{(n)}}/{\left(1+\rho_{e}^{(n)}\right)}+{1}/{\upsilon_{e}^{(n)}}+z_{2,k_{e}}.
\end{align*}

\begin{figure*}[!t] 
	\normalsize 
	\begin{align} 
		\xi_{i} \sqrt{V_{i}(\wko,\pmt)} &\leq   x_{2,i}^{(n+1)}-2\Re\left\{ {\sum}_{m=1}^M \pmb{f}_{i}^{(n+1)} e^{(j \theta_m)}\right\} +\left(e^{(j \pmt)}\right)^H \pmb{\varphi}_{2,i}^{(n+1)} e^{(j \pmt)}, i \in \{k,e\}. 
		\label{urllcbob2} \tag{26}\\
		x_{2,i}^{(n+1)}&\triangleq\frac{\xi_i{\left(\left(\upsilon^{(n+1)}_i\right)^2 \sqrt{V_i(\wko,\thetak)}+2\left(\upsilon^{(n+1)}_i\right)^2+2\sigma_{i}\rho_{i}^{(n+1)}-4\sigma_i \left(\upsilon^{(n+1)}_i\right)\right)}}{{\left(2\left(\upsilon^{(n+1)}_i\right)^2 \sqrt{V_i(\wko,\thetak)}\right)}}, \label{q2k} \tag{27}\\
		\pmb{f}_{i,j}^{(n+1)}(m) &\triangleq  {\xi_i}\bigg/{\left(\upsilon^{(n+1)}_i \sqrt{V_i(\wko,\thetak)}\right)}  \left(\wko_j \right)^H \pmb{h}_{i}^H(\thetak) \pmb{l}_i \Lambda_m {\pmb{L}}_{_{\text{{AR}}}} \wko_j ,m \in M, j\in K, j \neq k,\label{mkj1} \tag{28}\\
		x_{k}^{(n+1)} &\triangleq {x}_{1,k}^{(n+1)}+{x}_{1,k_{e}}^{(n+1)}+ x_{2,k}^{(n+1)}+ x_{2,k_{e}}^{(n+1)}+z_{1,ke}^{(n+1)}- \left(e^{j \thetak}\right)^H \pmb{\varphi} _{k}^{(n+1)}e^{j\btheta}-2\lambda_{\text{max}}\left(\pmb{\varphi}_{k}^{(n+1)}\right) M, \tag{30} \label{qk2}\\
		\pmb{y}^{(n+1)}_k(m) &\triangleq \pmb{y}^{(n+1)}_{k,k}(m)+\pmb{y}^{(n+1)}_{k_{e}}(m)+\pmb{f}^{(n+1)}_{k}(m)+\pmb{f}^{(n+1)}_{k_{e}}(m)+ {\sum}_{m=1}^{N}e^{-j \theta_m^{(n)}} \pmb{\varphi}_{k}^{(n+1)}(m,n)+\lambda_{\text{max}}\left(\pmb{\varphi}_{k}^{(n+1)}\right).\tag{31} \label{mk2}	
	\end{align} 
	\hrulefill 
\end{figure*}
Finally, by substitution  \eqref{SRFBRtot} in \eqref{probw}, and introducing an auxiliary variable $\mathcal{D}$ as the lower bound of the FBR-SR, we can recast problem ($\mathcal{P}1.1$) as:
\begin{subequations} \label{solw}
	\begin{alignat}{3} 
		&(\mathcal{P}1.3): && \max_{\Gamma, \pmw} \mathcal{D}, \label{solw-1} \\            
		&~\text{s.t.} &&  \mathcal{D}\leq \widetilde{\mathcal{S}}_{k}^{\mathcal{F}}(\pmw,\thetak),~\forall k, \label{solw-12} \\
		& && \eqref{P1-2},~\eqref{P1-4}, \label{solw-12x}
	\end{alignat}
\end{subequations}
problem ($\mathcal{P}1.3$) is a Semidefinite Programming problem (SDP), with its linearized objective function in \eqref{SRFBRtot}. This problem can be solved using standard solvers, e.g., the interior point method or the CVX toolbox \cite{grant2014cvx}. 
\subsubsection{Sub-Problem for Optimizing the PREs} \label{subsec2}
Likewise, given $\pmw$, we aim to find $\btheta^{(n+1)}$ such that, $\mathcal{S}_{k}^{\mathcal{F}}(\wko_k,\thetako) \geq \mathcal{S}_{k}^{\mathcal{F}}(\wko_k,\thetak)$. 
Similar to the previous section, the lower bound approximation of the user's SR can be obtained using the inequality \eqref{fund1} in Appendix \ref{AppA}. The user's data rate can be expressed as in \eqref{LBBob2x}, where 
\begin{alignat*}{2}
	&{x}_{1,k}^{(n+1)} &&\triangleq C_k(\wko,\thetak)-\gamma_k(\wko,\thetak)-\sigk {z}_{1,k}^{(n+1)},\\
	&{z}_{1,k}^{(n+1)} &&\triangleq{1}/{\rho_{k}^{(n+1)}}-{1}/{\upsilon_{_{k}}^{(n+1)}},\\
	&\pmb{y}^{(n+1)}_{k,k}(m) &&\triangleq \tilde{\pmb{y}}^{(n+1)}_{k,k}(m)/{\rho_{k}^{(n+1)}}, \\
	&\tilde{\pmb{y}}^{(n+1)}_{k,k}(m) &&\triangleq \left(\wko_k \right)^H\pmb{h}_{_{k}}^H(\thetak) \pmb{l}_k  \pmb{\Gamma}_m \pmb{L}_{\text{AR}}  \wko_k, \\
	&\pmb{\varphi}_{1,k}^{(n+1)} &&\triangleq -{z}_{1,k}^{(n+1)}{\sum}_{j=1}^{\mathcal{K}} \pmb{\varphi}_{k,j}^{(n+1)}, \\
	&\pmb{\varphi}_{k,j}^{(n+1)} &&\triangleq \left(\pmb{\Lambda}_{k,j}^{(n+1)}(m)\right)^* \pmb{\Lambda}_{k,j}^{(n+1)}(\dot{m}), m\in M, \dot{m} \in M, \\
	&\pmb{\Lambda}_{k,j}^{(n+1)}(m) &&\triangleq \pmb{l}_k \pmb{\Gamma}_m \pmb{L}_{_{\text{AR}}} \wko_j.
\end{alignat*}
\begin{algorithm} 
	\caption{Proposed Iterative Algorithm for Solving $(\mathcal{P}1)$} \label{alg1}
	\begin{algorithmic}[1]
		\State \textbf{Initialize:} $(\pmb{w}^{(1)}=\pmw^*,~\btheta^{(1)}=\pmt^*$, ~$\tau_{\max}$,~$\tau_{e_{\max}}$,~$t_{\max}$), convergence tolerance $\epsilon_t > 0$,  and Set $n=1$. 
		\State \textbf{Repeat} 
		\State Update $\wko$ by \eqref{solw}, and $\thetako$ by \eqref{solth};
		\State \textbf{if} $\frac{|\underset{k \in \mathcal{K}}{\min} \mathcal{S}_{k}^{\mathcal{F}}(\wko,\thetako)-\underset{k \in \mathcal{K}}{\min} \mathcal{S}_{k}^{\mathcal{F}}(\wk,\thetak)|}{\underset{k \in \mathcal{K}}{\min} \mathcal{S}_{k}^{\mathcal{F}}(\wk,\thetak)} \leq \epsilon_t$.
		\State \textbf{Then} $\thetak\leftarrow \thetako$, $\wk\leftarrow \wko$ and terminate. 
		\State \textbf{Otherwise} $n\leftarrow n+1$ and continue.
		\State \textbf{Output} $(\pmw^*=\wk_k, \pmt^*=\thetak)$.
	\end{algorithmic}
\end{algorithm}
Similarly, we can express Eve's eavesdropping rate as in \eqref{LBEve2x}, where, 
\begin{alignat*}{2}
	&x_{1,k_{e}}^{(n+1)} &&\triangleq \ln\left(\rho_{e}^{(n+1)}\right)-\rho_{e}^{(n+1)}-\ln\left(\upsilon_{e}^{(n+1)}\right)+1, \\
	&z_{1,k_{e}}^{(n+1)} &&\triangleq\left(-{\rho_{e}^{(n+1)}}/{1+\rho_{e}^{(n+1)}}-{1}/{\upsilon_{e}^{(n+1)}}\right)\sige, \\
	&{\pmb{y}}^{(n+1)}_{k_e}(m) &&\triangleq {\sum}_{j=1,j \neq k}^{\mathcal{K}} {\pmb{y}}^{(n+1)}_{j,k_{e}}(m), \\
	&{\pmb{y}}^{(n+1)}_{j,k_{e}}(m) &&\triangleq \left(\wko_k \right)^H\pmb{h}_{e}^H(\thetak) \pmb{l}_{e}  \pmb{\Gamma}_m {\pmb{L}}_{_{\text{{AR}}}}  \wko_k, \\
	&\pmb{\varphi} _{1,k_{e}}^{(n+1)} &&\triangleq -\dfrac{\rho_{e}^{(n+1)}{\sum}_{j=1}^{\mathcal{K}} {\pmb{\varphi}}_{k_{e},j}^{(n+1)} }{(1+\rho_{e}^{(n+1)})}-\frac{{\sum}_{j=1}^{\mathcal{K}} {\pmb{\varphi}}_{k_{e},j}^{(n+1)} }{(\sige+\upsilon_{e}^{(n+1)})}, \\
	&\pmb{\varphi}_{k_{e},j}^{(n+1)} &&\triangleq \left({\pmb{\Lambda}}_{k_{e},j}^{(n+1)}(m)\right)^* \pmb{\Lambda}_{k_{e},j}^{(n+1)}(\dot{m}), \\
	&\pmb{\Lambda}_{k_{e},j}^{(n+1)}(m) &&\triangleq \pmb{l}_e \pmb{\Gamma}_m {\pmb{L}}_{_{\text{{AR}}}} \wko_j.
\end{alignat*}

Next, we express the user-$k$'s and Eve dispersion factor. For $i \in \{k,e\}$, we can express the dispersion factors as in \eqref{urllcbob2},
where $x_{i,k}^{(n+1)}$ is expressed as in \eqref{q2k}, $\pmb{f}_{i,j}^{(n+1)}(m)$ is expressed as in \eqref{mkj1}, and 
\begin{align*}
	\pmb{f}_{i}^{(n+1)} &\triangleq{\sum}_{j=1, j \neq k}^\mathcal{K} \pmb{f}_{i,j}^{(n+1)}, \\
	\pmb{\varphi}_{2,i}^{(n+1)}&\triangleq{z}_{2,i}^{(n+1)} {\sum}_{j=1}^{\mathcal{K}} \pmb{\varphi}_{i,j}^{(n+1)}, \\
	{z}_{2,i}^{(n+1)}&\triangleq \dfrac{(\xi_{i} \rho^{(n+1)}_i)}{\left(\upsilon^{(n+1)}_i\right)^2 \sqrt{V_i(\wko,\thetak)}}.
\end{align*}
By substituting \eqref{LBBob2x}, \eqref{LBEve2x}, and \eqref{urllcbob2} in (\ref{Cstot2}), the lower bounding concave approximation of the FBR-SR can be expressed  as:
\setcounter{equation}{28}
\begin{align} \label{SRtheta}
	\widetilde{\mathcal{S}}_k^{\mathcal{F}}\left(\wko,\btheta\right)& \geq x_{k}^{(n+1)}+ 2{\sum}_{m=1}^{M}\Re\left\{\pmb{y}^{(n+1)}_{k}(m) e^{j \theta_m}\right\},
\end{align}
where $	x_{k}^{(n+1)}$ is expressed as in \eqref{qk2}, ${\pmb{y}}^{(n+1)}_k(m)$ is expressed as in \eqref{mk2}, and 
\begin{align*}
	\pmb{\varphi}_{k}^{(n+1)} &\triangleq \pmb{\varphi}_{1,k}^{(n+1)}+\pmb{\varphi}_{1,k_{e}}^{(n+1)}+\pmb{\varphi}_{k_{e},j}^{(n+1)}+\pmb{\varphi}_{2,k}^{(n+1)}+\pmb{\varphi}_{2,k_{e}}^{(n+1)}. 
\end{align*}
With (\ref{SRtheta}), we formulate the following optimization problem to obtain $\thetako$, \setcounter{equation}{31}
\begin{align}
    (\mathcal{P}1.4):~\underset{\btheta}{\max}~ \underset{k \in \mathcal{K}}{\min}~ \widetilde{\mathcal{S}}_k^{\mathcal{F}}\left(\wko,\thetak\right),~\text{s.t.}~\eqref{P1-3}.
\end{align}
To tackle problem ($\mathcal{P}1.4$) we define,  
\begin{align} \label{convx2}
	\theta_m^{(n+1),k} \triangleq 2\pi - \angle {\pmb{y}}^{(n+1)}_k(m), m=1,\dots,M,
\end{align}
then we find $\btheta^{(n+1)}$ that maximize the $\mbox{SR}(\wko,\pmt)$, and satisfies the UMC constraint \eqref{P1-3},
\begin{align} \label{solth}
	\thetako \triangleq \arg \underset{\pmt \in \left\{\btheta^{(n+1),k}, k=1,\dots,\mathcal{K}\right\}}{\max} \widetilde{\mathcal{S}}_{k}^{\mathcal{F}}(\wko,\pmt).
\end{align}
One can notice that by setting $V_i = 0$, the above FBR optimization problem $(\mathcal{P}1)$ is reduced to the LBR optimization problem, which can be solved similarly.
\begin{figure*}[!t] 
	\normalsize 
	\begin{align}
		&\mathbf{A}_k  \triangleq \mathbf{\Phi} \pmb{L}_{\text{AR}} \pmw_k \pmw_k^{(n),H} \pmb{L}_{\text{AR}}^H \mathbf{\Phi}^{(n),H}+ \mathbf{\Phi}^{(n)} \pmb{L}_{\text{AR}} \pmw_k^{(n)} \pmw_k^{H} \pmb{L}_{\text{AR}}^H \mathbf{\Phi}^{(n),H} - \mathbf{\Phi}^{(n)} \pmb{L}_{\text{AR}} \pmw_k^{(n)} \pmw_k^{(n),H} \pmb{L}_{\text{AR}}^H \mathbf{\Phi}^{(n),H} \label{LMI1-1}, \tag{40} \\
		 &\Delta\pmb{l}_{k} \mathbf{A}_k \Delta\pmb{l}_{k}^H + 2 \Re \{( \widehat{\pmb{l}}_{k}^H \mathbf{A}_{k}) \Delta\pmb{l}_{k}\}+ d_k \geq (2^{\beta_{k}}-1)\alpha_{k}, \forall \|\Delta\pmb{l}_k\|_2 \leq \Omega_k, \forall k. \tag{41} \label{Bobx1}
	\end{align}
	\hrulefill 
\end{figure*}

The procedure to solve problem ($\mathcal{P}1$) is described in Algorithm \ref{alg1}, which converges to a locally optimal solution of ($\mathcal{P}1$) as formally stated in the following theorem.
\begin{theorem} \label{theo1}
	The obtained solution by Algorithm \ref{alg1} is a locally optimal solution for problem $(\mathcal{P}1)$.
\end{theorem}
\textit{Proof}: See Appendix \ref{AppB}.
\subsubsection{Complexity Analysis}
The developed Algorithm \ref{alg1} is designed to tackle problem $(\mathcal{P}1)$ by decoupling the beamforming vector and the IRS's PREs within the objective function. The resulting problem $(\mathcal{P}1)$ is an SDP that can be solved by the interior point method\cite{grant2014cvx}. The algorithm's complexity can be estimated by its worst-case runtime and the number of decision variables \cite{labit2002sedumi}. 
Thus, in Algorithm \ref{alg1}, the computational complexity of obtaining $\pmw$ given $\btheta$ is $\mathcal{O} (N^3)$, and the computational complexity of obtaining $\btheta$ given $\pmw$ is $\mathcal{O}(M^{3}(M+1))$. One can notice that the algorithm complexity increases with $M$.
		\section{MAXMIN FBR-SR under Imperfect CSI } \label{sec4} %
        To deal with the imperfect CSI from the IRS to the users and Eve, we adopt the channel modeling in equation \eqref{ICSIcha} where only partial/imperfect CSI $\widehat{\pmb{l}}_{i}$ is available. First, we can cast the minimum SR maximization problem as follows: \setcounter{equation}{34}
\begin{subequations} \label{P6}
	\begin{alignat} {2}
		&(\mathcal{P}2): && \underset{\pmw,\btheta}{\max}  ~ \underset{k \in \mathcal{K}}{\min} ~ \widehat{\mathcal{S}}_{k}^{\mathcal{F}}(\pmw,\btheta), \label{P2-1} \\ 
		&~\text{s.t.} &&  {\sum}_{k = 1}^{\mathcal{K}}\Vert \pmw_k \Vert^2 \leq P, \label{P2-2} \\ 
		& && |e^{(j \btheta)}| = 1, \label{P2-3} \\
		& && \|\Delta\pmb{l}_i\|_2 \leq \Omega_{i}, i \in \{k,e\}. \label{P2-4}\\
		& && \tau_k\leq \tau_{\max}, ~\tau_e \leq \tau_{e_{\max}}, ~t_t \leq t_{\max}, \label{P2-5}
	\end{alignat}
\end{subequations}
where \eqref{P2-4} captures the channel estimation error. Unlike the traditional LBR case under imperfect CSI, problem $(\mathcal{P}2)$ is more challenging. First, this is because the additional FBR constraints, which are captured by the nonconvex dispersion factor expression \eqref{dispfa2}, are embedded within the objective function. In addition, the imperfect CSI condition \eqref{P2-4}, introduces uncertainty in the IRS-users/Eve's channel, making the decoding error and consequently the inverse Q-function evaluation unreliable. This sensitivity significantly complicates the objective function, besides its nonlinearity and nonconvexity.

To solve problem $(\mathcal{P}2)$, we first introduce slack variables to decompose the coupling of the beamforming vector and the IRS's PREs in the objective function to facilitate the AO method. Specifically, we first substitute \eqref{Cstot} into \eqref{P2-1}, and then we introduce slack variables $\mathcal{Z}$ as the FBR-SR's lower bound, ${\beta}_k$ represents the minimum users' rate, $\Upsilon_{k_e}$ represents the maximum eavesdropping rate of Eve, and $ \tilde{\beta}_i, i \in \{k,e\}$ represents the maximum dispersion factor for the user-$k$ and Eve, respectively. The problem $(\mathcal{P}2)$ can be recast as: 
\begin{subequations} \label{P6.1}
	\begin{alignat}{7} 
		&(\mathcal{P}2.1): && ~ \underset{\pmw,\btheta}{\max} ~ \mathcal{Z}, \label{P6-1} \\ 
		&~\text{s.t.} &&  \mathcal{Z} \leq {\beta}_k - \Upsilon_{k_e} - \tilde{\beta}_k - \tilde{\beta}_{k_e}, \forall k, \label{PwCSI-2x} \\ 
		& && {\beta}_k \leq \widehat{C}_k, \forall \|\Delta\pmb{l}_k\|_2 \leq \xi_{k}, \forall k, \label{P6-2} \\
		& && \Upsilon_{k_e} \geq \widehat{C}_e, \forall \|\Delta\pmb{l}_e\|_2 \leq \xi_{e}, \forall k, \label{P6-3} \\
		& && \tilde{\beta}_k \geq  \xi_k \sqrt{\widehat{V}_{k}}, \forall \|\Delta\pmb{l}_k\|_2 \leq \xi_{k}, \forall k, \label{P6-4}\\
		& && \tilde{\beta}_{k_e} \geq \xi_{e} \sqrt{\widehat{V}_{e}}, \forall \|\Delta\pmb{l}_e\|_2 \leq \xi_{e}, \forall k, \label{P7-5x} \\
		& && \eqref{P2-2},~\eqref{P2-3},~\eqref{P2-5}. \label{P7-6x}
	\end{alignat}
\end{subequations}
Problem $\mathcal{P}2.1$ is still nonconvex due to the coupling between $\pmw$ and $\btheta$ within the constraints \eqref{PwCSI-2x}, \eqref{P6-2}, \eqref{P6-3}, \eqref{P6-4}, and \eqref{P7-5x}. In addition, the presence of the nonlinear square-root terms of the dispersion factors in \eqref{P6-4} and \eqref{P7-5x} further increases the complexity of the problem. One can notice that the constraint \eqref{P2-4} in problem $(\mathcal{P}2)$ has been captured by the constraints \eqref{P6-2}, \eqref{P6-3}, \eqref{P6-4}, and \eqref{P7-5x} in problem $(\mathcal{P}2.1)$ (referring to the CSI factor in the channel definition \eqref{chaCSIdef}).  Next,  we linearize the square-root of the dispersion factors, and then we employ the SCA technique, the $\mathcal{S}$-procedure, and the first order Taylor approximation to transform the semi-infinite nonconvex constraints \eqref{P6-2}, \eqref{P6-3}, \eqref{P6-4}, and \eqref{P7-5x} to finite LMIs. 
Then, we leverage a PCCP algorithm to tackle the UMC \eqref{P2-3} \cite{lipp2016variations}. 
\subsubsection{Sub-Problem for Optimizing the Beamforming Vectors} \label{subsec3}
To linearize the semi-infinite inequalities in \eqref{P6-2}, we first substitute \eqref{ckedef} into  \eqref{P6-2}, hence, \eqref{P6-2} can be expressed as:
\begin{align} \label{P3CSI-BnLx2}
	2^{\beta_k}-1 \leq \dfrac{\left|(\pmb{l}_k \pmb{\Phi} \pmb{L}_{\text{AR}})\pmw_{k}\right|^2}{\left\|(\pmb{l}_k \pmb{\Phi} \pmb{L}_{\text{AR}})\pmW_{-k}\right\|^2 +\sigk},
\end{align}
where $\pmW_{-k} = [\pmw_1,\dots,\pmw_{k-1},\pmw_{k+1},\dots,\pmw_{\mathcal{K}}] \in \mathbb{C}^{M \times \mathcal{K}-1}$. By treating the interference plus noise signal as an auxiliary function $\pmb{\alpha} = [\alpha_1, \dots, \alpha_{K}]$, \eqref{P3CSI-BnLx2} can be expressed as: 
\begin{subequations}
	\begin{equation} \label{P3CSI-BLx3}
		\left|(\pmb{l}_k \pmb{\Phi} \pmb{L}_{\text{AR}})\pmw_{k}\right|^2 \geq (2^{\beta_k}-1)\alpha_{k}, \forall k,
	\end{equation}
	\begin{equation} \label{P3CSIn-BLx2}
		\left\|(\pmb{ u}_k \pmb{\Phi} \pmb{L}_{\text{AR}})\pmW_{-k}\right\|^2 +\sigk \leq \alpha_{k}, \forall k.
	\end{equation}
\end{subequations}
To circumvent the nonconvex semi-infinite inequalities in \eqref{P3CSI-BLx3}, we replace the left-hand side of  \eqref{P3CSI-BLx3} with their lower bounds using the following Lemma. 

\begin{mylem} \label{lem1_1}
	At iteration $(n)$, let $\wk$ and $\thetak$ be the optimal solution, then at the point $(\wk,\thetak)$ we can express the linear lower bound of  \eqref{P3CSI-BLx3} as:
	\begin{align} \label{ICSIeqw}
		\left|(\pmb{l}_k \pmb{\Phi} \pmb{L}_{\text{AR}})\pmw_k\right|^2 \triangleq \pmb{l}_k \mathbf{A}_k \pmb{l}_k^H, 
	\end{align}
	where $\mathbf{A}_k$, is given in \eqref{LMI1-1}.
\end{mylem}

\Prf Refer to Appendix \ref{apendixB}. 

Next, by substituting \eqref{ICSIeqw} in  \eqref{P3CSI-BLx3}, and using \eqref{ch2-h1} and Lemma \ref{lem1_1}, the inequality in \eqref{P3CSI-BLx3} is reformulated as in \eqref{Bobx1}, where $d_k \triangleq \widehat{\pmb{l}}_{k} \mathbf{X}_{k} \widehat{\pmb{l}}_{k}^H$.

To address the uncertainty of $\{\Delta\pmb{l}_k\}$ in \eqref{Bobx1}, we leverage the $\mathcal{S}$-procedure \cite{boyd1994linear}. 
\begin{mylem} \label{lem1_2}
	($\mathcal{S}$-procedure) For any Hermitian matrix $\mathbf{U}_i \in \mathbb{C}^{L \times L}$, vector  $\mathbf{u}_i \in \mathbb{C}^{L \times 1}$, and scalar $\text{u}_i$, for ${i}= 0,\dots,Q$. A quadratic function of a variable $x$ is defined as: \setcounter{equation}{41}
	\begin{align}
		f_{i}(x) \triangleq x^H \mathbf{U}_{i} x + 2\Re\{\mathbf{u}_{i}^H x\}+{\text{u}}_{i}.
	\end{align}
	The condition $f_{0}(x) \geq 0$ such that $f_{i}(x) \geq 0, ~ i = 1,\dots,Q$,  holds, if an only if there exists $\mathfrak{n}_{i} \geq 0, i= 0,\dots,Q$, such that,
	\vspace{-0.3cm}\begin{align} 
		\begin{bmatrix}
			\mathbf{U}_{0} & \mathbf{u}_{0} \\ \mathbf{u}_{0}^H & \text{u}_{0}	
		\end{bmatrix} - {\sum}_{i= 0}^{Q} \mathfrak{n}_{i} \begin{bmatrix}
			\mathbf{U}_{i} & \mathbf{u}_{i} \\ \mathbf{u}_{i}^H & \text{u}_{i}	
		\end{bmatrix} \succeq 0.
	\end{align} 
\end{mylem}

Using Lemma \ref{lem1_2}, we can transform \eqref{Bobx1} into its equivalent LMIs as:
\begin{align} \label{P3CSI-BLX3n2}
	\begin{bmatrix}
		\eta_{k} \mathbf{I}_{M}+ \mathbf{A}_k & (\widehat{\pmb{l}}_{k} \mathbf{A}_k)^H \\ (\widehat{\pmb{l}}_{k} \mathbf{A}_k) & d_k - (2^{\beta_k}-1)\alpha_{k} - \eta_{k} \Omega_{k}^2	
	\end{bmatrix} \succeq 0, \forall k,
\end{align}
where $\pmb{\eta} \triangleq [\eta_{1},\dots,\eta_{\mathcal{K}}]^T \geq 0$ are slack variables. Even though \eqref{P3CSI-BLx3} has been transformed into an LMI form, it is still nonconvex due to the nonconvexity nature of the term $2^{\beta_k} \alpha_{k}$. For that, we adopt the SCA technique \cite{9197675} to convert the nonconvex constraint \eqref{P3CSI-BLx3} to a convex approximation expression. Specifically, performing the first order Taylor approximation, the term $2^{\beta_k} \alpha_{k}$ can be upper bounded by:
\begin{align} \label{upbeta}
	(\alpha_{k}(2^{\beta_k}))^{ub} \triangleq \left((\beta_k-\beta_k^{(n)})(\alpha_{k}^{(n)}) \ln2 + \alpha_{k}\right)2^{\beta_k^{(n)}},
\end{align}
where $\beta_k^{(n)}, \alpha_{k}^{(n)}$ are the value of the variables $\beta_k, \alpha_{k}$ at iteration $(n)$ in the SCA-based algorithm, respectively. Lastly, substituting \eqref{upbeta} in \eqref{P3CSI-BLX3n2}, the LMIs in \eqref{P3CSI-BLX3n2} can be expressed as:
\begin{align} \label{P3CSI-BLX3n3}
	\begin{bmatrix}
		\eta_{k} \mathbf{I}_{M}+ \mathbf{A}_k & (\widehat{\pmb{l}}_{k} \mathbf{A}_k)^H \\ (\widehat{\pmb{l}}_{k} \mathbf{A}_k) & d_k - (\alpha_{k}(2^{\beta_k}))^{ub}+\alpha_{k} - \eta_{k} \Omega_{k}^2	
	\end{bmatrix} \succeq 0, \forall k.
\end{align} 

Next, we tackle  \eqref{P3CSIn-BLx2} using Schur's complement \cite{boyd2004convex}.
\begin{mylem} \label{lem1_3}
	(Schur's complement) For given matrices $\mathbf{U} \succeq 0$, $\mathbf{Y}$, and $\mathbf{Z}$, let a Hermitian matrix $\mathbf{X}$ be defined as: 
	\begin{align} 
		\mathbf{X}\triangleq
		\begin{bmatrix}
			\mathbf{Z} & \mathbf{Y}^H \\ \mathbf{Y} & \mathbf{U}
		\end{bmatrix}.
	\end{align} 
\end{mylem}
Then, $\mathbf{X} \succeq 0$ if and only if $\Delta{\mathbf{U}} \succeq 0$, where $\Delta{\mathbf{U}}$ is the Schur's complement define as $\Delta{\mathbf{U}} \triangleq \mathbf{Z}-\mathbf{Y}^H\mathbf{U}^{-1}\mathbf{Y}$.

Using Lemma \eqref{lem1_3}, we can equivalently recast \eqref{P3CSIn-BLx2} as:
\begin{align} \label{BobLMIx2n}
	\begin{bmatrix}
		\alpha_{k} & \mathbf{t}_k^H \\ \mathbf{t}_{k}^H & \mathbf{I}_{K-1}
	\end{bmatrix} \succeq 0, \forall \|\Delta\pmb{l}_k\|_2 \leq \Omega_{k}, \forall k,
\end{align}
where $\mathbf{t}_{k} \triangleq \left((\widehat{\pmb{l}}_{k}^H \mathbf{\Phi} \pmb{L}_{\text{AR}})\mathbf{W}_{-k}\right)^H$.

Next, we use the Nemirovski’s Lemma \cite{1369660} to further handle \eqref{BobLMIx2n}.
\begin{mylem} \label{lem1_4}
	(Nemirovski’s Lemma) For any Hermitian matrix $\mathbf{A}$, matrices $\mathbf{B}$, $\mathbf{C}$, and $\mathbf{X}$, and scalar $t$, the following LMI holds,
	\begin{align}
		\mathbf{A} \succeq \mathbf{B}^H \mathbf{X} \mathbf{C}+\mathbf{C^H \mathbf{X}^H} \mathbf{B}, ~\text{for} \|\mathbf{X}\| \leq t,
	\end{align}
	if and only if 
	\begin{align} 
		\begin{bmatrix}
			\mathbf{A}- a\mathbf{C}^H \mathbf{C} & -t \mathbf{B}^H \\ -t\mathbf{B} & a \mathbf{I}
		\end{bmatrix} \succeq 0,
	\end{align} 
\end{mylem}
where $a$ is a non-negative real number.

Using Lemma \eqref{lem1_4} and introducing the slack variable  $\pmb{\varkappa} \triangleq [\varkappa_1, \dots, \varkappa_k] \geq 0$, \eqref{BobLMIx2n} is recast as: 
\begin{align} \label{P3CSI-BLX4n}
	\begin{bmatrix}
		Q_{11,k}& \widehat{t}^H_k & \mathbf{0}_{1 \times M }  \\
		\widehat{t}_k & \mathbf{I}_{K-1} & \Omega_{k} (\mathbf{\Phi} \pmb{L}_{\text{AR}} \pmW_{-k})^H  \\  \mathbf{0}_{M \times 1 } & \Omega_{k} (\mathbf{\Phi} \pmb{L}_{\text{AR}} \pmW_{-k}) & \varkappa_k \mathbf{I}_{M} 
	\end{bmatrix} \succeq 0, \forall k,
\end{align}
where $Q_{11,k} \triangleq \alpha_k - \sigk - \varkappa_k$, and $\widehat{t}_k \triangleq \left((\widehat{\pmb{l}}_{k} \mathbf{\Phi} \pmb{L}_{\text{AR}})\mathbf{W}_{-k}\right)^H$.
Next, we tackle \eqref{P6-3} by firstly substituting \eqref{ckedef} in  \eqref{P6-3}, and then treat the interference plus noise signal in  \eqref{P6-3} as an auxiliary function $\pmb{\alpha}_{e_k}\triangleq [\alpha_{e_1},\dots,\alpha_{e_k}]$. We hence can recast \eqref{P6-3} as
\begin{subequations}
	\begin{alignat} {2}
		&\left|(\pmb{l}_{e} \mathbf{\Phi} \pmb{L}_{\text{AR}})\pmw_{k}\right|^2 \leq (2^{\Upsilon_{k_e}}-1)\alpha_{ke}, \forall k,\label{EveLMI1} \\
		&\left\|(\pmb{l}_{e} \mathbf{\Phi} \pmb{L}_{\text{AR}})\pmW_{-k}\right\|^2 +\sige \geq \alpha_{ke}, \forall k. \label{EveLMI2}
	\end{alignat}
\end{subequations}

To tackle the uncertainties of $\{\Delta\pmb{l}_e\}$ in the constraints \eqref{EveLMI1} and \eqref{EveLMI2}, we use a similar approach as in \eqref{P3CSIn-BLx2}. Hence, the equivalent LMIs of \eqref{EveLMI1} and \eqref{EveLMI2} are, respectively, given by,
\begin{alignat}{2}
	\begin{bmatrix}
		C_{e,k}& \widehat{b}_{ke}^H & \mathbf{0}_{1 \times M }  \\
		\widehat{b}_{ke} & 1 & \Omega_e (\mathbf{\Phi} \pmb{L}_{\text{AR}} \pmw_k)^H  \\  \mathbf{0}_{M \times 1 } & \Omega_{e} (\mathbf{\Phi} \pmb{L}_{\text{AR}} \pmw_k) & \vartheta_{k} \mathbf{I}_{M} 
	\end{bmatrix} \succeq 0, \forall k, \label{EveLMI11} \\
	\begin{bmatrix}
		B_{11,k}& -\widehat{b}_{ke}^H & \mathbf{0}_{1 \times M }  \\
		-\widehat{b}_{ke} & \mathbf{I}_{K-1} & -\Omega_e (\mathbf{\Phi} \pmb{L}_{\text{AR}} \pmW_{-k})^H  \\  \mathbf{0}_{M \times 1 } & -\Omega_{e} (\mathbf{\Phi} \pmb{L}_{\text{AR}} \pmW_{-k}) & \varrho_{k} \mathbf{I}_{M} 
	\end{bmatrix} \succeq 0, \forall k, \label{EveLMI22}
\end{alignat}
where   $\pmb{\vartheta} \triangleq [\vartheta_{1}, \dots, \vartheta_{K}] \geq 0$ and $\pmb{\varrho} \triangleq [\varrho_{1}, \dots, \varrho_{K}] \geq 0$ are a slack variables, and,
\begin{align*}
	C_{e,k} & \triangleq (\alpha_{ke}(2^{\Upsilon_{k_e}}))^{ub}-\alpha_{ke}-\vartheta_{k}, \\
	(\alpha_{ke}(2^{\Upsilon_{k_e}}))^{ub} & \triangleq ((\Upsilon_{k_e}-\Upsilon_{k_e}^{(n)})(\alpha_{ke}^{(n)}) \ln2 + \alpha_{ke})2^{\Upsilon_{k_e}^{(n)}}, \\
	\widehat{b}_{ke} &\triangleq ((\widehat{\pmb{l}}_{e}^H \mathbf{\Phi} \pmb{L}_{\text{AR}})\mathbf{W}_{-k})^H, \\
	B_{11,k} &\triangleq \alpha_{ke} - \sige - \varrho_{k}.
\end{align*}
Next, we show the steps to derive the LMI of equations \eqref{P6-4} and \eqref{P7-5x}. Using \eqref{dispfa2},  the inequalities \eqref{xy} and \eqref{xt} in Appendix \ref{AppA}, and by defining $x\triangleq V_i$, $\mathbf{A} \triangleq \rho_{i}$, and $B \triangleq \tau_i$, the dispersion factors for $i \in \{k,e\}$ can be expressed as:
\begin{align*} 
	\xi_{i}\sqrt{V_{i}}& \leq \frac{\xi_{i} \sqrt{V_{i}^{(n)}}}{2}+ \frac{\xi_{i} |(\pmb{l}_i \pmb{\Phi} \pmb{L}_{\text{AR}})\pmw_k|^2}{\sqrt{V_{i}^{(n)}}\left(\|(\pmb{l}_i \pmb{\Phi} \pmb{L}_{\text{AR}}) \pmW_{k}\|_2^2+\sigma_i\right)},
\end{align*}
where $\mathbf{W}_k \triangleq [\pmw_{1},\dots,\pmw_{\mathcal{K}}] \in \mathbb{C}^{M \times \mathcal{K}}$. 
Hence, we can express \eqref{P6-4} and \eqref{P7-5x} as, 
\begin{align} \label{P3CSI-BFx2x}
	\frac{(\tilde{\beta}_{i}-\mathfrak{L})}{\mathfrak{Y}} \geq    \frac{|(\pmb{l}_{i} \pmb{\Phi} \pmb{L}_{\text{AR}})\pmw_{k}|^2}{\|(\pmb{l}_{i} \pmb{\Phi} \pmb{L}_{\text{AR}}) \pmW_{k}\|_2^2+\sigma_{i}},
\end{align}
where $\mathfrak{L} \triangleq {\xi_{i} \sqrt{V_{i}^{(n)}}}/{2}$ and $\mathfrak{Y} \triangleq {\xi_{i}}/{\sqrt{V_{i}^{(n)}}}$. 

Similar to \eqref{P3CSI-BnLx2}, by treating the interference plus noise signal as an auxiliary function $\pmb{\zeta} \triangleq [\zeta_{1}, \dots, \zeta_{K}]$, \eqref{P3CSI-BFx2x} can be expressed as:
\begin{subequations}
	\begin{alignat} {2}
		&\left|(\pmb{l}_i \pmb{\Phi} \pmb{L}_{\text{AR}})\pmw_{k}\right|^2  \leq \zeta_{i}{(\tilde{\beta}_{i}-\mathfrak{L})}/{\mathfrak{Y}} , \forall k, i \in \{k,e\}, \label{BobFLMI1}\\
		&\left\|(\pmb{ u}_{i} \pmb{\Phi} \pmb{L}_{\text{AR}})\pmW_{k}\right\|^2 +\sigma_{i} \geq \zeta_{k}, \forall k, i \in \{k,e\}. \label{BobFLMI2}
	\end{alignat}
\end{subequations}
Similar to \eqref{EveLMI22}, we can express \eqref{BobFLMI2} by its equivalent LMI  as: 
\begin{align} \label{BobFLMI22}
	\begin{bmatrix}
		\widetilde{C}_{11,i}& -\tilde{c}^H_i & \mathbf{0}_{1 \times M }  \\
		-\tilde{c}_i & \mathbf{I}_{K} & -\Omega_i (\mathbf{\Phi} \pmb{L}_{\text{AR}}) \pmW_{k})^H  \\  \mathbf{0}_{M \times 1 } & -\Omega_{i} (\mathbf{\Phi} \pmb{L}_{\text{AR}}) \pmW_{k}) & \mathfrak{l}_{i} \mathbf{I}_{M} 
	\end{bmatrix} \succeq 0, 
\end{align}
where $\widetilde{C}_{11,i} \triangleq \zeta_{k} - \sigk - \mathfrak{u}_{i}$,  $\pmb{\mathfrak{l}} \triangleq [\mathfrak{l}_{1}, \mathfrak{u}_{2}, \dots, \mathfrak{l}_{\mathcal{K}}] \geq 0$ is a slack variable, and $\tilde{c}_i \triangleq ((\widehat{\pmb{l}}_{i}^H \mathbf{\Phi} \pmb{L}_{\text{AR}}))\mathbf{W}_{k})^H$.

Next, we can express \eqref{BobFLMI1} as the following LMI:
\begin{align} \label{BobFLMI11x}
	\begin{bmatrix}
		D_{11,i}& \tilde{d}_{i}^H & \mathbf{0}_{1 \times M }  \\
		\tilde{d}_{i} & 1 & \Omega_e (\mathbf{\Phi} \pmb{L}_{\text{AR}}) \pmw_k)^H  \\  \mathbf{0}_{M \times 1 } & \Omega_{e} (\mathbf{\Phi} \pmb{L}_{\text{AR}}) \pmw_k) & \tilde{\mathfrak{l}}_{i} \mathbf{I}_{M} 
	\end{bmatrix} \succeq 0,  
\end{align}
\begin{algorithm}
\caption{PCCP Algorithm for solving $(\mathcal{P}2.3)$} \label{alg3}
\begin{algorithmic}[1]
    \State \textbf{Initialize:} $(\pmb{w}^{(1)},\btheta^{(1)})$, penalty factor $a_{\max}$, scaling $\nu \!\ge\! 1$, tolerances $(\epsilon_{t_1},\epsilon_{t_2})$, $n_{\max}$, set $n=1$. 
    \Repeat
        \State Solve $(\mathcal{P}2.3)$ to obtain $\thetako$.
        \If{$\|e^{j\btheta^{(n)}}-e^{j\btheta^{(n-1)}}\|_1 \le \epsilon_{t_2}$ and $C \le \epsilon_{t_1}$}
            \State \textbf{break}
        \Else
            \State Update $a^{(n+1)}=\min\{\nu a^{(n)},a_{\max}\}$, set $n\!=\!n+1$.
        \EndIf
    \Until{$n>n_{\max}$}
    \State \textbf{Output:} feasible solution $\btheta^*=\thetak$.
\end{algorithmic}
\end{algorithm}
where $D_{11,i} \triangleq \frac{\tilde{\beta}_{i}\zeta_{i}}{\mathfrak{Y}}-\frac{\mathfrak{L}\zeta_{i}}{\mathfrak{Y}} -\tilde{\mathfrak{l}}_{i}$,  $\tilde{\mathfrak{l}} \triangleq [\tilde{\mathfrak{l}}_{1}, \dots, \tilde{\mathfrak{l}}_{\mathcal{K}}] \geq 0$ is a slack variable, and $\tilde{d}_{i} \triangleq ((\widehat{\pmb{l}}_{e} \mathbf{\Phi} \pmb{L}_{\text{AR}})\pmw_{k})^H$.

However, \eqref{BobFLMI11x} is still nonconvex due to the nonconvex nature of the term $\tilde{\beta} \zeta_{k}$. Similar to \eqref{P3CSI-BLX3n2}, we employ the SCA technique to tackle the nonconvex expression as: 
\begin{align} \label{upperterm2}
	\left(\tilde{\beta} \zeta_{i}\right)^{ub} \triangleq \tilde{\beta}^{(n)} \zeta_{i} + (\tilde{\beta}-\tilde{\beta}^{(n)})\zeta_{i}^{(n)}
\end{align}
where $\tilde{\beta}^{(n)}$, $\zeta_{i}^{(n)}$ are the value of the variables $\tilde{\beta}_{i}$, $\zeta_{i}$ at iteration $(n)$ in the SCA-based algorithm, respectively. Be substituting \eqref{upperterm2} in \eqref{BobFLMI11x}, we can express \eqref{BobFLMI11x} as:
\begin{align} \label{BobFLMI11}
	\begin{bmatrix}
		\widetilde{D}_{i}& \tilde{b}^H_{ke} & \mathbf{0}_{1 \times M }  \\
		\tilde{d}_{i} & 1 & \Omega_{e} (\mathbf{\Phi} \pmb{L}_{\text{AR}}) \pmw_k)^H  \\  \mathbf{0}_{M \times 1 } & \Omega_{e} (\mathbf{\Phi} \pmb{L}_{\text{AR}}) \pmw_k) & \tilde{\mathfrak{l}}_{i} \mathbf{I}_{M} 
	\end{bmatrix} \succeq 0,  j\neq k,
\end{align}
where $\widetilde{D}_{i} \triangleq \frac{\left(\tilde{\beta} \zeta_{i}\right)^{ub}}{\mathfrak{Y}}-\frac{\mathfrak{L}\zeta_{i}}{\mathfrak{Y}} -\tilde{\mathfrak{l}}_{i}$.

Eventually, reformulating problem $(\mathcal{P}2.1)$ with \eqref{P3CSI-BLX3n3}, \eqref{P3CSI-BLX4n}, \eqref{EveLMI11}, \eqref{EveLMI22}, \eqref{BobFLMI22}, and \eqref{BobFLMI11} yields the following optimization problem:
\begin{subequations}\label{P6.3}
	\begin{alignat}{2}
		&(\mathcal{P}2.2):  && ~\underset{\pmw, \btheta}{\max} ~ \mathcal{Z} \label{P6.3-1} \\
		&~ \text{s.t.}       && \mathcal{Z} \leq {\beta}_k - \Upsilon_{k_e} - \tilde{\beta}_k - \tilde{\beta}_{k_e}, \forall k, \\
		& && \eqref{P3CSI-BLX3n3}, \eqref{P3CSI-BLX4n}, \eqref{EveLMI11}, \eqref{EveLMI22}, \eqref{BobFLMI22}, \eqref{BobFLMI11}, i \in \{k,e\}, \\
		& && \pmb{\eta}\geq 0, \pmb{\varkappa} \geq 0, \pmb{\vartheta} \geq 0, \pmb{\varrho} \geq 0, \mathfrak{l} \geq 0, \tilde{\mathfrak{l}} \geq 0, \\
		& && \eqref{P2-2},~\eqref{P2-5}.
	\end{alignat}
\end{subequations}
At this point, all nonlinear constraints \eqref{P6-2}, \eqref{P6-3}, \eqref{P6-4}, and \eqref{P7-5x} have been linearized as LMIs \eqref{P3CSI-BLX3n3}, \eqref{P3CSI-BLX4n}, \eqref{EveLMI11}, \eqref{EveLMI22}, \eqref{BobFLMI22}, and \eqref{BobFLMI11}; thus, problem ($\mathcal{P}2.2$) is an SDP that can be efficiently solved using standard solvers, e.g., the interior point method or the CVX toolbox \cite{grant2014cvx}. 
\subsubsection{Sub-Problem for Optimizing the PREs} \label{subsec4}
For a given $\pmw$, we aim to find $\thetako$ such that, $\widehat{\mathcal{S}}^{\mathcal{F}}_{k}(\wko_k,\thetako) > \widehat{\mathcal{S}}^{\mathcal{F}}_{k}(\wko_k,\thetak)$. Similar to the analysis in the previous section, we can express $\eqref{P6-2}$ and $\eqref{P6-3}$ with their equivalent LMIs \eqref{P3CSI-BLX3n3}, \eqref{P3CSI-BLX4n}, \eqref{EveLMI11} and \eqref{EveLMI22}, respectively. To tackle the UMC \eqref{P2-3}, we leverage the PCCP method \cite{9525400}. The PCCP's main idea is to add slack variables to relax the problem so that if the UMC is violated, we then can penalize the sum of violations. The converged PCCP solution is an approximate first-order optimal solution of the original problem \cite{9525400}. 
\begin{algorithm}
\caption{Proposed AO Algorithm for Solving Problem $(\mathcal{P}2)$} \label{alg4}
\begin{algorithmic}[1]
    \State \textbf{Initialize:} $(\pmb{w}^{(1)}=\pmw^{*},~\btheta^{(1)}=\btheta^{*})$, set $n=1$, and tolerance $\epsilon_t>0$.
    \Repeat
        \State Update $\wko$ by solving $(\mathcal{P}2.2)$ and $\thetako$ by solving $(\mathcal{P}2.3)$.
        \State $\Delta \!=\!\frac{|\min_{k\in\mathcal{K}}\mathcal{S}_{k}^{\mathcal{F}}(\wko,\thetako)-\min_{k\in\mathcal{K}}\mathcal{S}_{k}^{\mathcal{F}}(\wk,\thetak)|}{\min_{k\in\mathcal{K}}\mathcal{S}_{k}^{\mathcal{F}}(\wk,\thetak)}$.
        \State Update $\wk\!\leftarrow\!\wko,~\thetak\!\leftarrow\!\thetako$, and set $n\!\leftarrow\!n+1$.
    \Until{$\Delta \leq \epsilon_t$}
    \State \textbf{Output:} $(\pmw^{*}=\wk,~\btheta^{*}=\thetak)$.
\end{algorithmic}
\end{algorithm}

To apply the PCCP method, we introduce an auxiliary variable set $\mathbf{q} \triangleq \{q_{m}|m \in M\}$ satisfying $q_m \triangleq |e^{(j \theta_m)}|^* |e^{(j \theta_m)}|$. Then, \eqref{P2-3} can be expressed as $q_m \leq |e^{(j \theta_m)}|^* |e^{(j \theta_m)}| \leq q_m $. The nonconvex part $q_m \leq |e^{(j \theta_m)}|^* |e^{(j \theta_m)}|$ can be approximated by $q_m \leq 2 \Re \{|e^{(j \theta_m)}|^* |e^{(j \theta^{(n)}_n)}| - |e^{(j \theta^{(n)}_n)}|^* |e^{(j \theta^{(n)}_n)}|\}$. Following the PCCP framework, we penalize the objective function \eqref{P6-1}, hence, problem  $(\mathcal{P}2.1)$ can be recast as:
\begin{subequations} \label{P3CSIthBL}
	\begin{alignat} {2}
		&(\mathcal{P}2.3): && \underset{\btheta}{\max}  ~ \mathcal{Z} - a\mathcal{T}, \label{P3CSIthBL-1} \\
		&~ \text{s.t.}  && \eqref{P3CSI-BLX3n3}, \eqref{P3CSI-BLX4n}, \eqref{EveLMI11}, \eqref{EveLMI22}, \eqref{BobFLMI22}, \eqref{BobFLMI11}, \\
		& && |e^{(j \theta_m)}|^* |e^{(j \theta_m)}| \leq q_m + t_m, \label{P3CSIthBL-BLx1} \\
		& && q_m - \hat{t}_m \leq 2 \Re \left\{|e^{(j \theta_m)}|^* |e^{(j \theta^{(n)}_n)}|\right\}  \nonumber \\ 
		& &&- |e^{(j \theta^{(n)}_n)}|^* |e^{(j \theta^{(n)}_n)}|, \label{P3CSIthBL-BLx2} \\
		& &&  q_m \geq 0, \forall m \in M, \label{P3CSIthBL-BLx3}
	\end{alignat}
\end{subequations}
where $\mathcal{T} \triangleq {\sum}_{m=1}^N t_m+\hat{t}_m$ is the penalty term, $\mathbf{c} \triangleq \{c_n,\hat{c}_n\}$ is the slack variable imposed over the modulus constraint \eqref{P2-3}, and $a$ is the regularization factor. The regularization factor is imposed to control the UMC constraint \eqref{P2-3} by scaling the penalty term $Q$.

Problem $(\mathcal{P}2.3)$ is an SDP that can be solved by the CVX toolbox. Unlike the conventional SDR method to tackle UMC, the PCCP method, summarized in Algorithm \ref{alg3}, is guaranteed to find a feasible point for problem $(\mathcal{P}2.3)$ \cite{9774882}.  
The following points are helpful for the numerical implementation of the Algorithm \ref{alg3}:
\begin{enumerate}[label=(\alph*)]
	\item We invoke $a_{\text{max}}$ to avoid numerical complications if $a$ grows too large\cite{9505311};
	\item The stopping criteria $\mathcal{T}\leq \epsilon_{t_1}$ guarantees the UMC \eqref{P2-3} when $\epsilon_{t_1}$ is small \cite{9180053};
	\item The convergence of Algorithm \ref{alg3} is controlled by $\left\|e^{(j \btheta^{(n+1)})}-e^{(j \btheta^{(n)})}\right\|_1 \leq \epsilon_{t_2}$.
\end{enumerate}

Finally, the pseudo code of the AO algorithm to solve problem ($\mathcal{P}2$) is shown in Algorithm \ref{alg4}. This algorithm converges to a locally optimal solution of ($\mathcal{P}2$), which can be proved in the following theorem.
\begin{theorem}
	Algorithm \ref{alg4} generates a locally optimal solution for problem $(\mathcal{P}2)$.
\end{theorem}
\textit{Proof}: See Appendix \ref{AppD}.
\subsubsection{Complexity Analysis}
Since convex problems {$(\mathcal{P}2.2)$} and {$(\mathcal{P}2.3)$} involve linear constraints and LMIs {$\eqref{P3CSI-BLX3n3}, \eqref{P3CSI-BLX4n}, \eqref{EveLMI11}, \eqref{EveLMI22}, \eqref{BobFLMI22}$ and $\eqref{BobFLMI11},$} they can be solved by the interior point method\cite{grant2014cvx}. The algorithm's complexity is defined by its worst-case runtime and the number of operations \cite{9180053}. For problem {$(\mathcal{P}2.2)$}, the number of variables is ${a}\triangleq 2M$, the size of \eqref{P3CSI-BLX3n3} is ${b_1}\triangleq NM+N+1$, the size of \eqref{P3CSI-BLX4n}, and \eqref{EveLMI22} is ${b_2}\triangleq 2N+1$, and the size of  \eqref{EveLMI11} is ${b_3}\triangleq 2N+1$. Thus, the complexity of solving problems {$(\mathcal{P}2.2)$} and {$(\mathcal{P}2.3)$}, respectively,
\begin{align*}
	\mathcal{O}_{\pmw} &\triangleq \mathcal{O}\left\{\sqrt{{c_1}+2} (a)(a^2+a {c_2}+{c_3}+(a+1)^2)\right\}, \\
	\mathcal{O}_{\pmb{\Phi}} &\triangleq \mathcal{O}\left\{\sqrt{{c_1}+4M} (2M)(4M^2+2M {c_2}+{c_3}+4MN)\right\}, 
\end{align*}
where 
\vspace{-0.3cm}\begin{align*}
	{c_1} & \triangleq {\sum}_{k}^{\mathcal{K}}b_1(k-1)+b_2(2k+1)+b_3(3k-2), \\
	{c_2} & \triangleq {\sum}_{k}^{\mathcal{K}}b_1^2(k-1)+b_2^2(2k+1)+b_3^2(3k-2), \\
	{c_3} & \triangleq {\sum}_{k}^{\mathcal{K}}b_1^3(k-1)+b_2^3(2k+1)+b_3^3(3k-2), \\
\end{align*}

	\section{Numerical Results} \label{sec5}
	\begin{table}
		\centering
		\caption{Numerical Parameters}
		\begin{tabularx}{0.5\textwidth} {
				| >{\raggedright\arraybackslash}X
				| >{\centering\arraybackslash}X
				| >{\raggedleft\arraybackslash}X | }
			\hline
			Parameter & Numerical Value \\
			\hline
			Number of IRS elements ($M$)  & 16 \\
			\hline
			Noise power density ($\sigma_i$) & -174 dBm/Hz\\
			\hline
			Alice Transmission power ($P$) & 20 dBm\\
			\hline
			Antennas' Gain $(G_{\text{\text{A}}})$ and $(G_{\text{IRS}})$ & 5 dBi \\
			\hline
			The convergence tolerances $\epsilon_t, \epsilon_{t_1}, \epsilon_{t_1}$ & $10^{-3}$  \\
			\hline
			Sim. initial settings & $\alpha_{k}^{(n)}= 1$, $\alpha_{ke}^{(n)}= 1$, $a^{(1)}=10$, $a_{\text{max}}=30$, $\delta_{k}=\delta_{e}=0.02$.\\
			\hline
		\end{tabularx}
		\label{T2}
	\end{table}
	In this section, we perform an extensive simulation to evaluate the performance of the proposed approach. The results are obtained using MATLAB and CVX toolboxes.  In this setup, Alice is located at $(15,0,15)$, the IRS is located at $(0,25,40)$, and the users are randomly distributed to the right of Alice over a $(60m \times 60m)$ area. The Eve (e.g., a compromised legitimate user) is randomly located in $(100m \times 100m)$ outside the users' area. Note that if Eve is too close to one of the users, it is mainly impossible to guarantee the positive SR for all the users \cite{6772207}. In this case, other methods, e.g., encryption or friendly jamming, can provide users' secrecy \cite{9402750}. 
	
	The Alice-to-IRS direct path loss factor is ${\aleph_{\text{{AR}}}} \triangleq  \pmb{G}_{\text{\text{A}}}+\pmb{G}_{\text{IRS}}-35.9-22 \log_{10}(d_{_{\text{{AR}}}})$ in dB, where $d_{_{\text{{AR}}}}$ is the distance between Alice and the IRS in meters, while $\pmb{G}_{\text{\text{A}}}$ is Alice antenna gain, and $\pmb{G}_{\text{IRS}}$ is the IRS elements' antenna gain \cite{BOL19}. The path loss factor from the IRS to the users and the Eve is ${\aleph_{\text{R}i}} \triangleq  \pmb{G}_{\text{IRS}}-33.05-30 \log_{10}(d_{\text{R}i})$ dB, for $i\in\{k,e\}$, $d_{\text{R}i}$ is the distance between the IRS and the users and Eve in meters. 
	The spatial correlation matrix is $[\pmb{R}_{Ri}]_{l,\bar{l}} \triangleq  \exp{(j\pi (l-\bar{l}) \sin \hat{\vartheta} \sin \hat{\aleph})}$ for $i\in \{k,e\}$, where $\hat{\aleph}$ is the elevation angle and $\hat{\vartheta}$ is the azimuth angle \cite{kammoun2020asymptotic}. The elements of the Alice-to-IRS channel are generated by $[\pmb{G}_{_{\text{{AR}}}}]_{a,b} \triangleq  \exp({j\pi \left((b-1) \sin\overline{\Theta}_b \sin \overline{\vartheta}_b+(a-1) \sin(\theta_m) \sin(\vartheta_b)\right)})$, where ${\Theta}_n \in (0,2 \pi)$, ${\vartheta}_n \in (0,2 \pi)$, and $\overline{\Theta}_n \triangleq  \pi-\theta_m$, and $\overline{\vartheta}_n \triangleq  \pi+\vartheta_n$ \cite{kammoun2020asymptotic}. The small-scale fading channel gain $\widehat{\pmb{l}}_{i}$ for $ i \in \{k,e\}$ is modeled as a Rician fading channel with K-factor$=3$.  
	
	We define the CSI error bounds as $\Omega_{i} \triangleq \delta_i \|\widehat{\pmb{l}}_i\|_2, \forall k$, where $\delta_i \in [0,1), i\in \{k,e\}$ is the relative amount of CSI uncertainty. When $\delta = 0$, Alice can obtain the perfect CSI of the IRS to users/Eve reflected channel. In the FBR case, we set $\tau_{\max}=\tau_{e_{\max}}=10^{-5}$. The packet length $N_t$ is defined by the transmission duration $t_t$ and the bandwidth $\mathcal{B}$, as previously defined $N_t \triangleq \mathcal{B} t_t$. Hence, we set the transmission duration as $t_t=0.1$ ms, which is suitable for FBR transmission \cite{ben18}. The choice of $0.1$ ms end-to-end delay ensures having a quasi-static channel during FBR communication \cite{sh17cross}. The bandwidth $\mathcal{B}$ is set at $1$ MHz. Unless stated otherwise, the simulation parameters are defined in Table \ref{T2}. We multiply the results by $\log_2(e)$ to convert them to bps/Hz. Lastly, for comparison purposes, we compare our Max-Min algorithm against the SSR maximization algorithms. Due to the page limitation, the steps to maximize SSR has been omitted, however, the SSR maximization can be tackled by setting the objective function in problem ($\mathcal{P}1$) and problem ($\mathcal{P}2$) to $\underset{\pmw,\pmt}{\max} \sum_{k}^{\mathcal{K}} \mathcal{S}_{k}^{\mathcal{F}}(\pmw,\pmt)$.
	\begin{figure}[t!]
		\centering
		\includegraphics[width=0.36\textwidth]{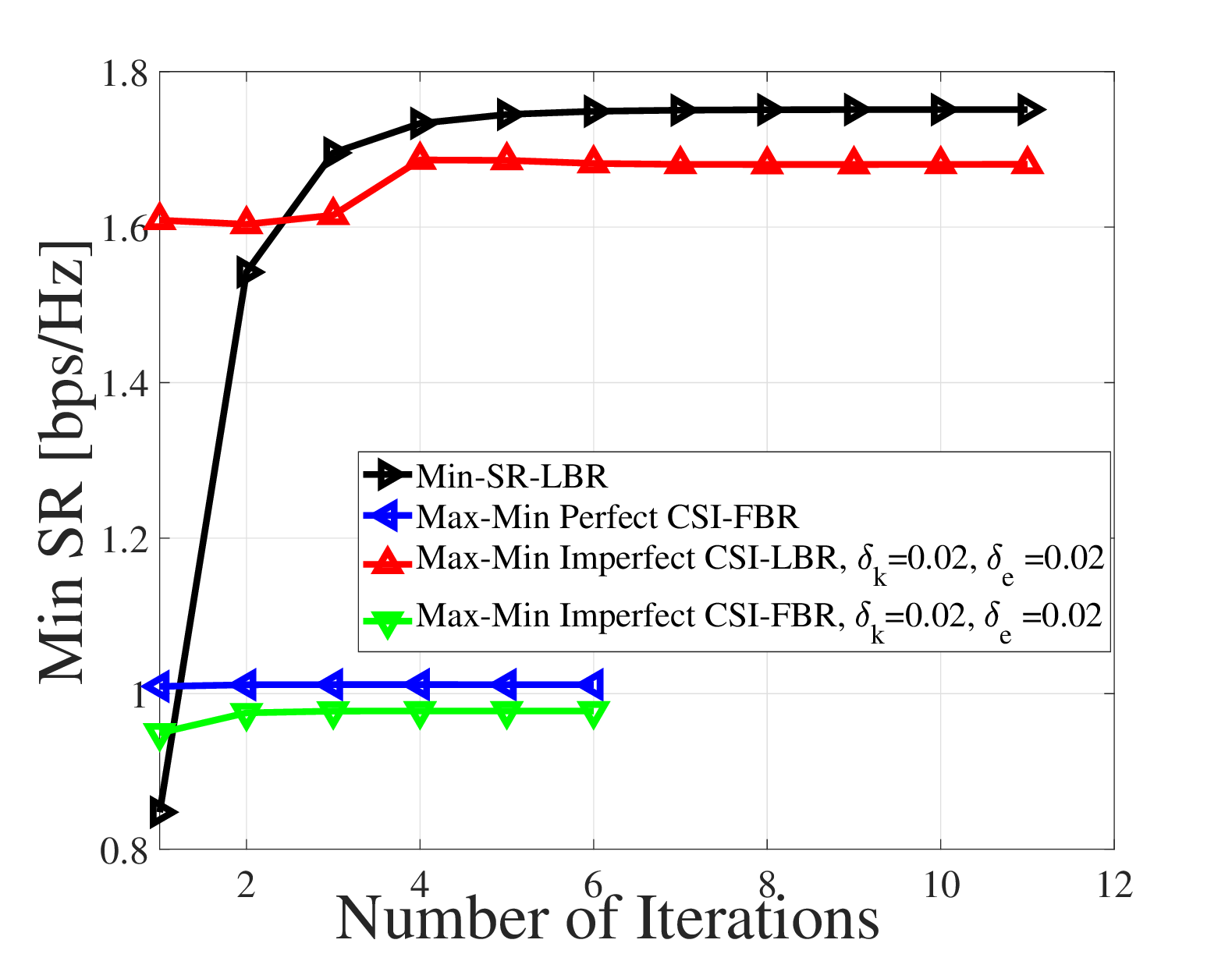}
		\caption{Convergence rate of the LBR/FBR Max-Min algorithm under perfect/Imperfect CSI with $N=10$, $\mathcal{K}=6$, $M=16$.}
		\label{Fig2}
	\end{figure}
	
	Fig.~\ref{Fig2}  illustrates the convergence rate of the Max-Min algorithms under perfect CSI and Imperfect CSI for the FBR and the LBR systems.  All algorithms achieve convergence within a few numbers of iterations. It can be noticed that the FBR needs fewer iterations to converge since the local optimal solution obtained in the LBR case is used as the initial feasible point. Such choice highlights the importance of choosing the best initial feasible points in the FBR case. 
		
\begin{figure}[t!]
		\centering
		\includegraphics[width=0.36\textwidth]{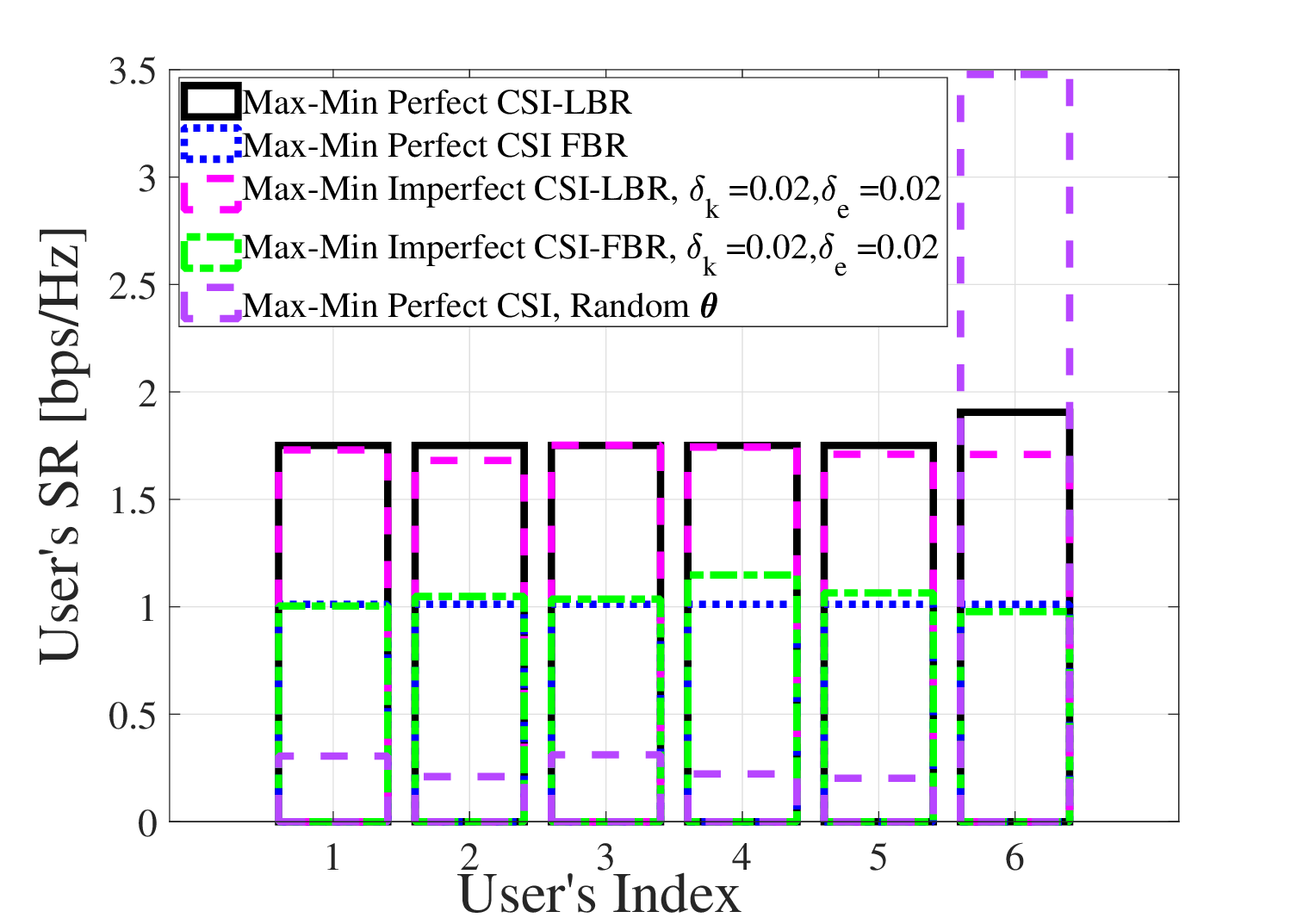}
		\caption{Users' SR with $N=10$, $\mathcal{K}=6$, $M=16$.}
		\label{Fig4}
	\end{figure}

	Fig. \ref{Fig4} plots the users' SR distribution with $N=10$, $\mathcal{K}=6$, and $M=16$ under the Max-Min algorithms. The Max-Min algorithms achieve almost the same SR for all the users when the IRS's PREs are optimized. However, when the IRS's PREs are not practically optimized, the algorithms fail to achieve secrecy fairness among the users. The results demonstrate the importance of properly optimizing the IRS's PREs. The proposed algorithm for the FBR case achieves secrecy and SR fairness among the users. The results demonstrate the robustness of the proposed algorithms to achieve secrecy for all the users, even under the FBR constraints, namely, the transmission duration and the latency. 
	\begin{figure}[t!] 
		\centering
		\includegraphics[width=0.36\textwidth]{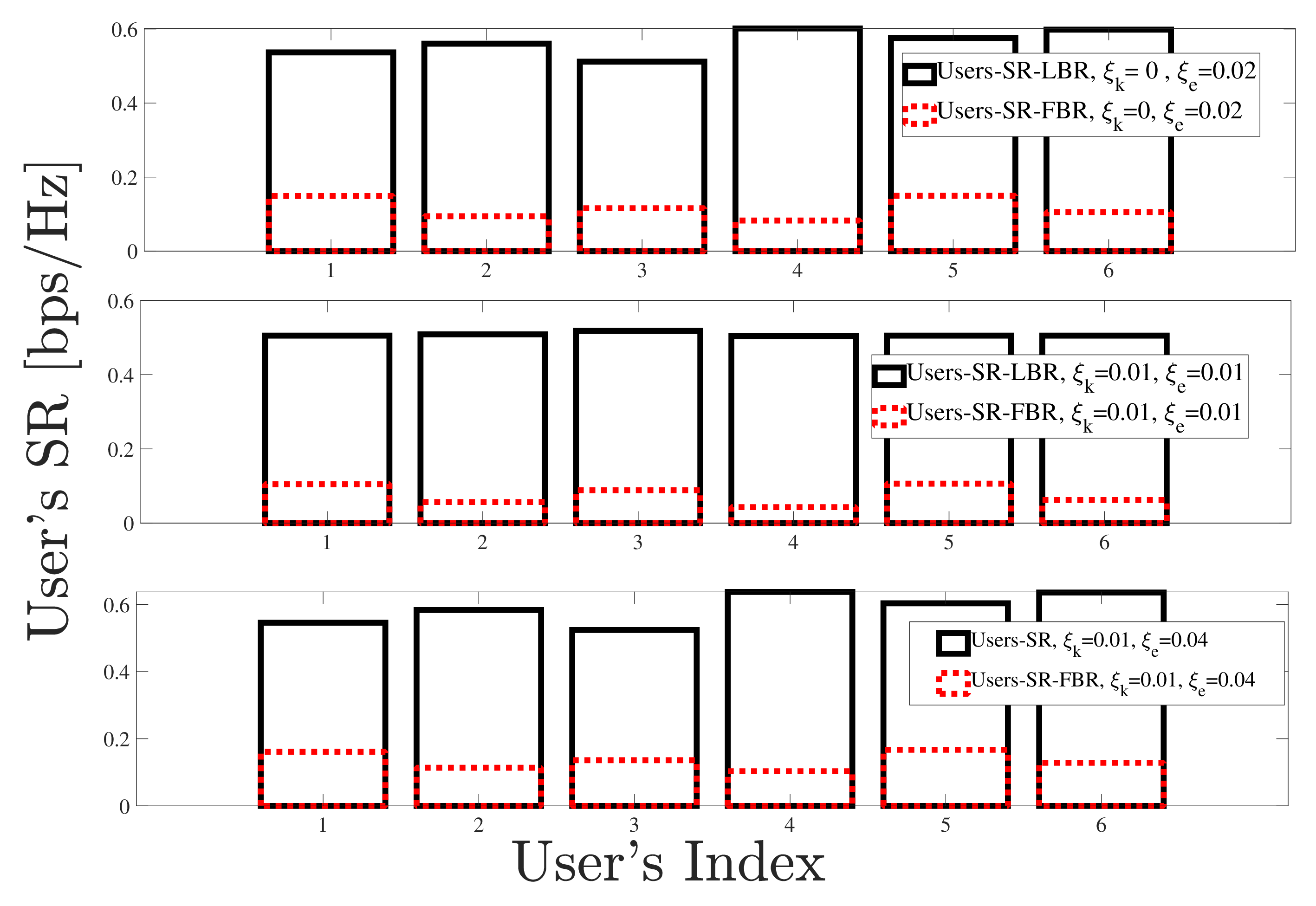}
		\caption{Users' SR with with $N=10, \mathcal{K}=6, M=16$.}
		\label{Fig8}
	\end{figure}
	
	Fig. \ref{Fig8} portrays the users' SR under different values of the IRS to the users/Eve's imperfect CSI. The Max-Min algorithm achieves secrecy fairness in the LBR/FBR cases, even when Eve's reflected channel has higher uncertainty than the users' reflected channel. The results show that the proposed algorithm is able to achieve secrecy guarantee even under a high level of uncertainty, which demonstrates the robustness of the proposed algorithm.
	\begin{figure}[t!] 
		\centering
		\includegraphics[width=0.36\textwidth]{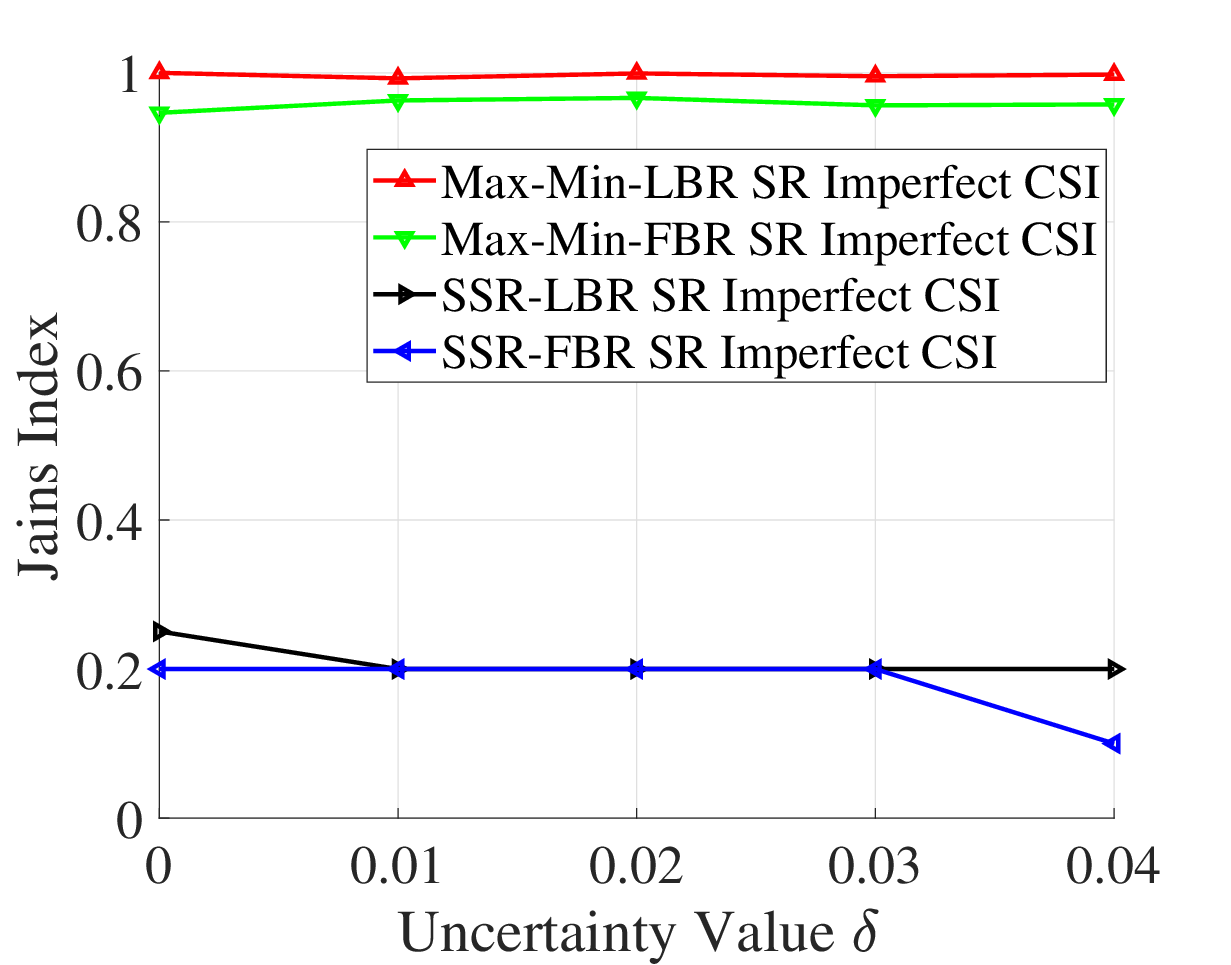}
		\caption{Jain's Index Vs. the uncertainty level $\delta_k=\delta_e$ with $N=10, \mathcal{K}=5, M=16$.}
		\label{Fig9}
	\end{figure}
	
	One way to evaluate the degree of fairness in the proposed algorithms is by using the Jain's index. Jain's index is defined as, $\text{Jain's Index}= \frac{\left(\sum_{k=1}^{\mathcal{K}}	\mathcal{S}_{\mathcal{K}}(\pmw,\btheta)\right)^2}{\mathcal{K} \sum_{k=1}^{\mathcal{K}}	(\mathcal{S}_{k}(\pmw,\btheta))^2}$, and it is bounded in $[1/\mathcal{K},1]$, where the higher value indicates a better fairness \cite{jain1984quantitative}. Fig. \ref{Fig9} shows Jain's index against the relative amount of CSI uncertainty $\delta$. The Max-Min algorithm achieves almost one in the LBR/FBR cases, while the SSR counterpart has nearly zero. The results demonstrate the Max-Min algorithm's robustness in achieving secrecy fairness among all the users, even in the FBR environment. 
	\begin{figure*}[!t] 
		\normalsize 
		\begin{align}
			&\ln\left|I_n+[\pmb{\Lambda}]^2(\pmb{\digamma})^{-1}\right|\geq \ln\left|I_n+[\hat{\Lambda}]^2(\hat{\digamma})^{-1}\right|
			-\la [\hat{\Lambda}]^2(\hat{\digamma})^{-1}\ra+2\Re\{\la \hat{\Lambda}^H(\hat{\digamma})^{-1}\mathbf{A}\ra\}
			-\la (\hat{\digamma})^{-1}-(\hat{\digamma}+[\hat{\Lambda}]^2)^{-1},[\mathbf{A}]^2+\pmb{\digamma}\ra , \tag{71} \label{fund1}  \\
			&\ln(1+\sum_{i=1}^{l} |{z}_i|^2) \geq \ln(1+\sum_{i=1}^{l} |\bar{z}_i|^2)-\sum_{i=1}^{l} |\bar{z}_i|^2 +\sum_{i=1}^{l} 2 \Re\{\bar{z}_i^* {z}_i\} - \frac{\sum_{i=1}^{l} |\bar{z}_i|^2 \left(1+\sum_{i=1}^{l} |{z}_i|^2\right)}{1+\sum_{i=1}^{l} |\bar{z}_i|^2}.  \tag{72} \label{fund2}
		\end{align} 
		\hrulefill 
	\end{figure*}

\begin{figure}[t!] 
		\centering
		\includegraphics[width=0.36\textwidth]{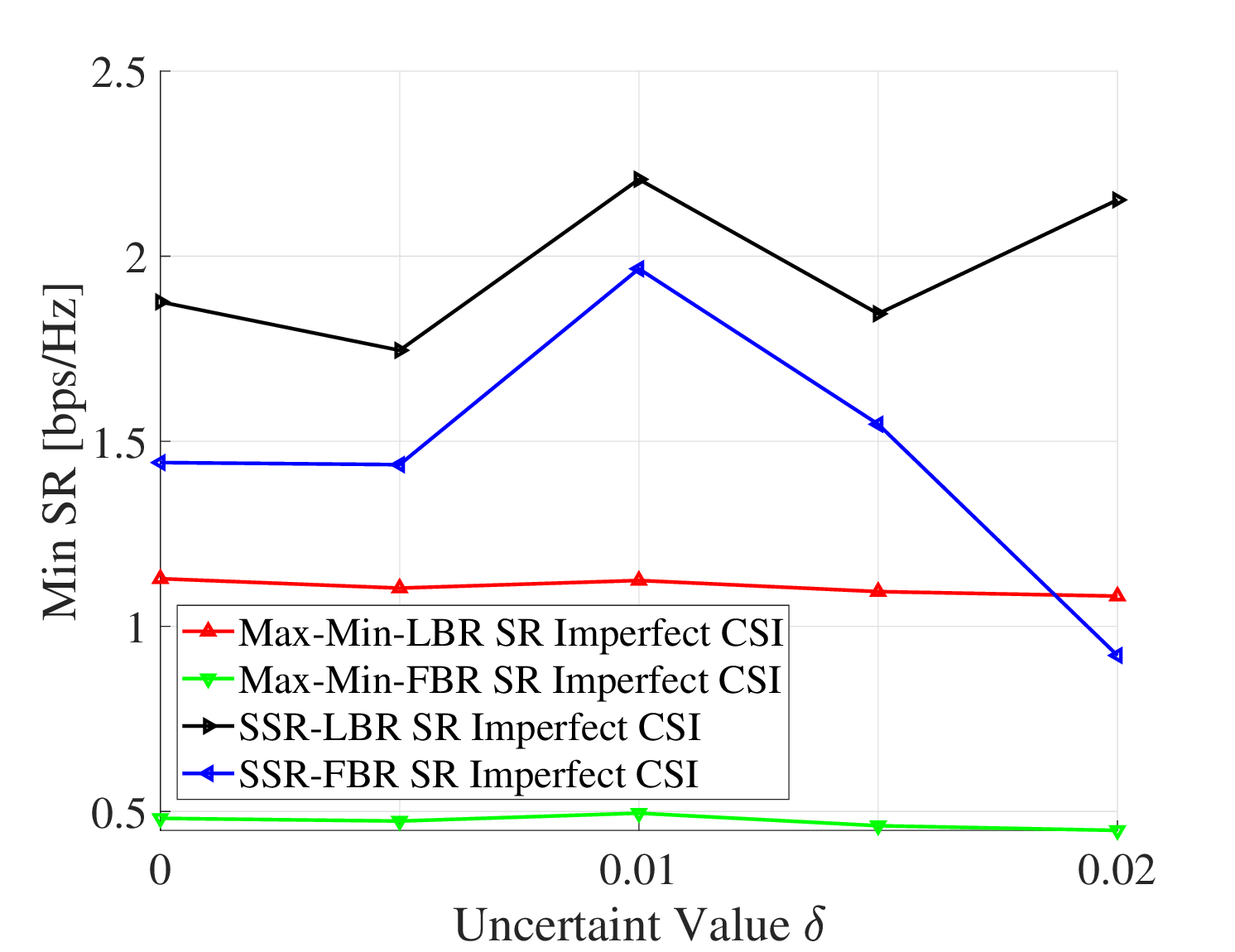}
		\caption{Minimum SR Vs. the uncertainty level $\delta_k=\delta_e$ with $N=10,\mathcal{K}=5, M=16$.}
		\label{Fig12}
	\end{figure}

	Fig. \ref{Fig12} illustrates the minimum SR among all users against the relative amount of CSI uncertainty $\delta$. The Max-Min algorithm achieves secrecy for all the users. Even under high uncertainty, it achieves almost the same minimum  SR as the full certainty case. The FBR case demonstrates results similar to those of the LBR case. As expected, the SSR-LBR/SSR-FBR algorithms achieve a higher minimum user SR since the SSR algorithm favors most of the transmission power towards the user with a better channel while discarding other users.
	\begin{figure}[t!] 
		\centering
		\includegraphics[width=0.36\textwidth]{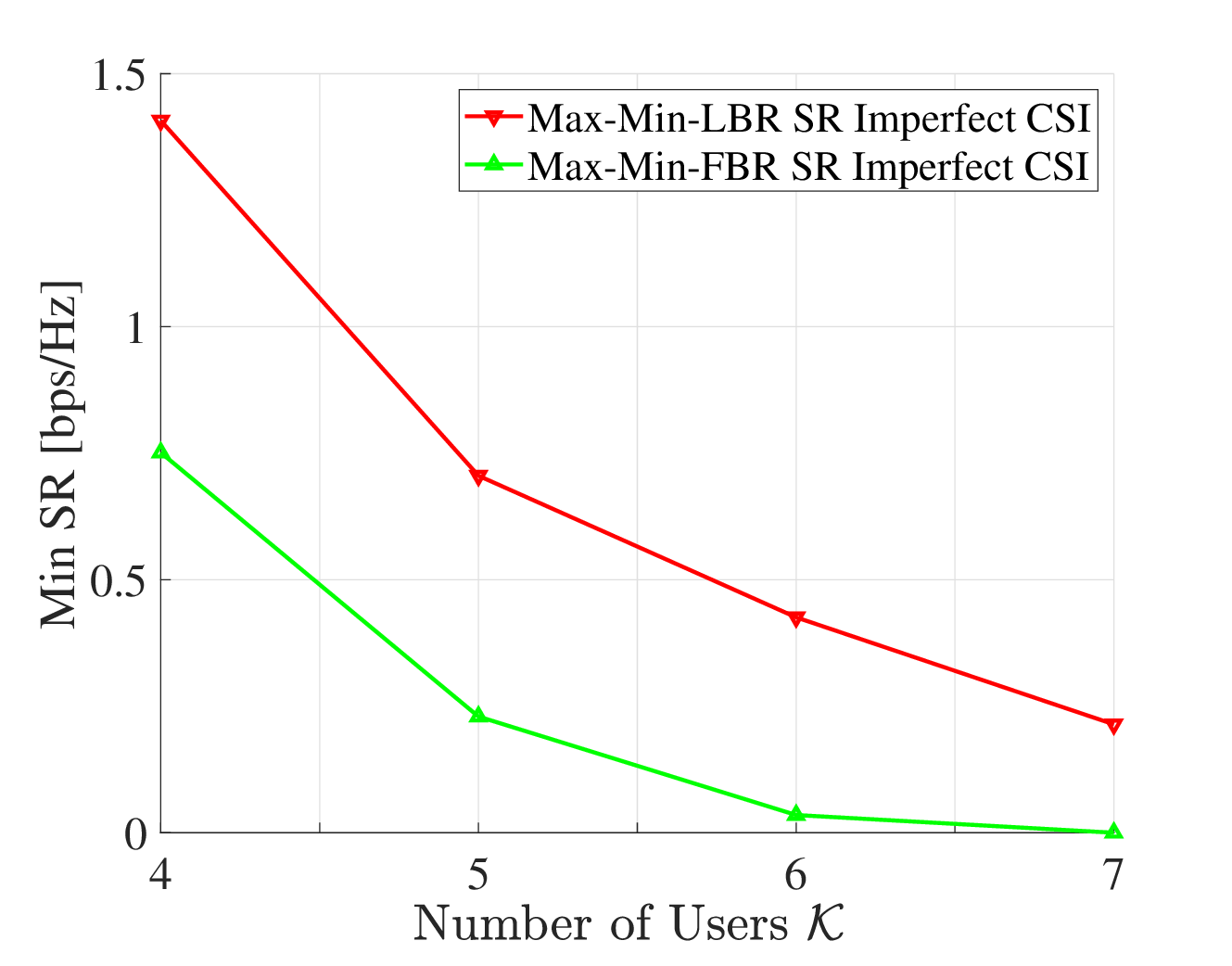}
		\caption{Minimum SR Vs. the number of users $\mathcal{K}$ with $N=10, M=16$, $\delta_{k}=\delta_{e}=0.02$.}
		\label{Fig10}
	\end{figure}
	
	Fig. \ref{Fig10} depicts the minimum SR against the number of users $\mathcal{K}$. It can be seen that the minimum SR algorithm decreases with $\mathcal{K}$. This is expected as the multi-user interference increases with $\mathcal{K}$. The Max-Min FBR case provides secrecy in all cases; however, as $\mathcal{K}$ increases, the minimum users' SR is approaching zero, which can be seen when $\mathcal{K}=7$. The results show that obtaining secrecy for all the users in the FBR case is more challenging due to the FBR constraints.
	\begin{figure}[t!] 
		\centering
		\includegraphics[width=0.36\textwidth]{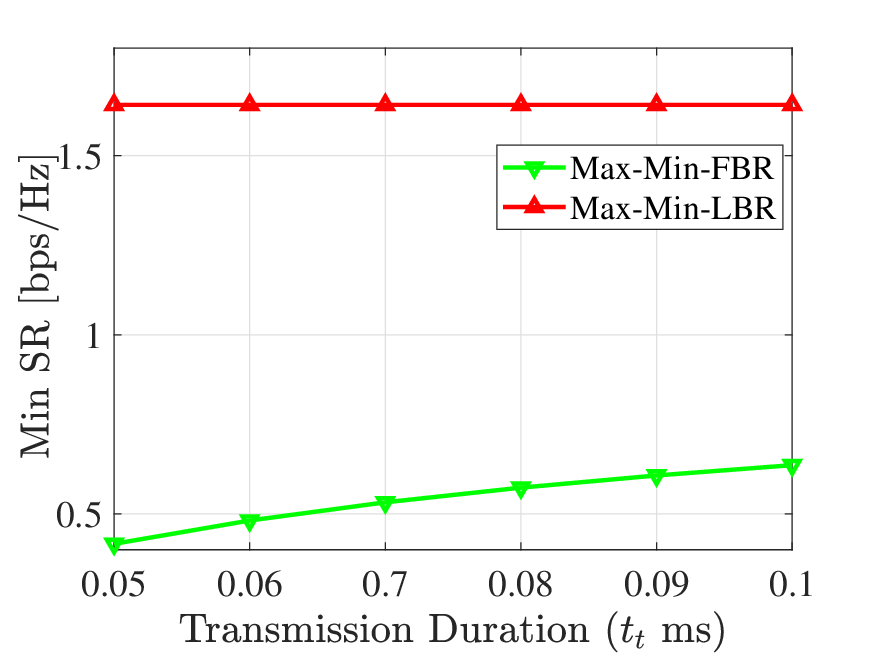}
		\caption{user's min SR Vs. Transmission Duration $t_t$.}
		\label{Fig11}
	\end{figure}
	
	Lastly, the minimum SR is illustrated against the transmission duration $t_t$ in Fig. \ref{Fig11}. The Max-Min algorithm guaratnees secure communication for all the users even at low transmission duration ($10^{-5}$ms). It can be noticed that the achievable FBR's SR is capped by its LBR counterpart since the transmission duration is considered to be infinity in the latter.

	\section{Conclusions} \label{conc}
	In this paper, we proposed a framework to achieve secure communications for all the users under the FBR constraints for IoT setting in URLLC applications aided by IRS. Specifically, through linearization and different nonconvex optimization techniques, we designed computationally efficient algorithm to maximize the minimum SR among all users under both perfect and imperfect CSI from IRS to the users and the eavesdropper.  Extensive simulations results showed that the Max-Min algorithm can provide secure communications for all the users under FBR constraints even with only imperfect CSI. Note that if Eve is too close to one user, it is infeasible to guarantee secrecy for all the users and other solutions like cryptography-based or friendly jamming-based methods should be in place. In the future, one can consider a multi-hop scenario with the joint coding over multiple IRSs, similar to the cooperative MIMO setting \cite{vmimo}.
	\begin{appendices}
		\section{Proof of theorem 1} \label{AppB}
		First we prove that the sequence $\mathcal{S}^{\mathcal{F}}_{k}(\wko,\thetako)$ is non decreasing for all $k$, i.e.  $\mathcal{S}^{\mathcal{F}}_{k}(\wko,\thetako) \geq \mathcal{S}^{\mathcal{F}}_{k}(\wk,\thetak)$ for all $n >0$. To that end, with the aid of the CVX solver, we obtain $\wko$ by solving problem ($\mathcal{P}1.3$). CVX solver is guaranteed to find the optimal solution. Thus, we have $\mathcal{S}^{\mathcal{F}}_{k}(\wko,\pmt) > \mathcal{S}^{\mathcal{F}}_{k}(\wk,\btheta)$ for all $k$. Next, from ($\mathcal{P}2$), we obtain $\thetako$ by solving  \eqref{convx2}, where we have $\mathcal{S}^{\mathcal{F}}_{k}(\pmw,\thetako) > \mathcal{S}^{\mathcal{F}}_{k}(\pmw,\thetak)$ as stated before. 
		
		Hence, by combining the solutions obtained by solving problem ($\mathcal{P}1.3$) and problem ($\mathcal{P}1.4$), we obtain: 
		\begin{align} \label{convxtot}
			\mathcal{S}^{\mathcal{F}}_{k}(\wko,\thetako) > \mathcal{S}^{\mathcal{F}}_{k}(\wk,\thetak),
		\end{align}
		the optimal sequence $\{(\wko,\thetako)\}$ thus converge to a point $\{({\pmw}^*,{\btheta}^*)\}$ which is the solution obtained from solving ($\mathcal{P}1.3$) and ($\mathcal{P}1.4$). 
		
		Next, we prove that the converged point $\hat{X}^* \triangleq \{\pmw^*,\btheta^*\}$ is a locally optimal solution of problem $(\mathcal{P}1)$. For that, we show that the converged point satisfies the Karush-Kuhn-Tucker (KKT) condition of the problem. The KKT condition for problem $(\mathcal{P}1.3)$ is satisfied at $\btheta^*$. Let $\pmb{Y}(\btheta)$ be the objective function of $	(\mathcal{P}2)$ and $\pmb{T}({\pmb{X}})=[\pmb{T}_1({\pmb{X}}), \pmb{T}_2({\pmb{X}}), \dots, \pmb{T}_I({\pmb{X}})]$ be the set of constraint of problem $(\mathcal{P}2)$. Then, we can write: 
		\begin{align} \label{con1}
			&\nabla_{\btheta^*}\pmb{Y}({\pmb{X}}^*)+ \pmb{Z}^T \nabla_{\btheta^*}\pmb{T}({\pmb{X}}^*) =0, \\
			& z_i \geq 0, z_i \pmb{T}({\pmb{X}}^*)=0, \forall i. \nonumber
		\end{align}
		where $\nabla_s$ is the gradient with respect to $s$, and $\pmb{Z} = [z_1,z_2,\dots,z_I]$ is the optimal Lagrangian variable set. Similarly, the obtained $(\mathcal{P}2)$ solution is locally optimal. Hence, its KKT is satisfied with respect to $\pmb{\text{W}}=\pmw_k^*$, which is: 
		\begin{align} \label{con2}
			&\nabla_{\pmb{\text{W}}^*}\pmb{Y}({\pmb{X}}^*)+ Z^T \nabla_{\pmb{\text{W}}^*}\pmb{T}({\pmb{X}}^*) =0, \\
			& z_i \geq 0, z_i \pmb{T}({\pmb{X}}^*)=0, \forall i. \nonumber
		\end{align}
		Combining \eqref{con1} and \eqref{con2}, we get:
		\begin{align}
			&\nabla_{{\pmb{X}}^*}\pmb{Y}({\pmb{X}}^*)+ Z^T \nabla_{{\pmb{X}}^*}\pmb{T}({\pmb{X}}^*) =0, \\
			& z_i \geq 0, z_i \pmb{T}({\pmb{X}}^*)=0, \forall i, \nonumber
		\end{align}
		which is the KKT condition for $(\mathcal{P}1)$ in \eqref{P1-1}, i.e., $\hat{\pmb{X}}^*$ is the local optimal point for \eqref{P1-1}. $\blacksquare$
		
		\section{Proof of Lemma 1} \label{apendixB}
		Let $q$ be a scalar complex variable, and $q^{(n)}$ is the fixed point obtained at iteration $(n)$, then the following inequality holds \cite{9110587}:
		\begin{align}
			|q|^2 \geq q^{*,(n)} q +q^{*} q^{(n)} - q^{*,(n)} q^{(n)}.
		\end{align}
		By replacing $q$ with $({\pmb{l}}_k \pmb{\Phi} \pmb{L}_{\text{AR}})\pmw_k$ we obtain \eqref{ICSIeqw}. Thus, this concludes the proof. $\blacksquare$

		\section{Proof of theorem 2} \label{AppD}
		Similar to the proof of Theorem \ref{theo1}, it can be shown the sequence $\mathcal{S}^{\mathcal{F}}_{k}(\wko,\thetako)$ is non-decreasing for all $k$, i.e.  $\mathcal{S}^{\mathcal{F}}_{k}(\wko,\thetako) \geq \mathcal{S}^{\mathcal{F}}_{k}(\wk,\thetak)$ for all $n >0$, and the converged point is a locally optimal solution for problems. $\blacksquare$.
		\section{Inequalities} \label{AppA}
		Here, we adopt the inequality (48) in\cite{TTN16}. Specifically, for any $\mathbf{A}$ and $\pmb{\digamma}$ with $\hat{\Lambda}$ and $\hat{\digamma}$ are fixed point, (\ref{fund1}) holds.
		
		Next, we adopt Lemma (2) in \cite{niu2022joint}. Specifically, for any ${z}_i$ with $i=1,\dots,l$ and $\bar{z}_i$ is a fixed point, inequality (\ref{fund2}) holds.

		Lastly, the concave logarithmic function can be written as \cite[Eq. 15]{niu2022joint}:
		\begin{align} \label{fund3}
			-\ln(1+\pmb{A}) \geq -\ln(1+\bar{A}) - \frac{1+\pmb{A}}{1+\bar{A}} +1.
		\end{align}
		The following inequality follows from the concavity of the function $\sqrt{x}$ \cite{abughalwa2022finite}:
		\begin{equation}\label{xy}
			\sqrt{x}\leq {\sqrt{\bar{x}}}/{2}\left(1+{x}/{\bar{x}}\right),~\forall x>0, \bar{x}>0.
		\end{equation}
		Lastly, for a any vector $\mathbf{A} \in\mathbb{C}^n$, $\bar{\mathbf{A}}\in\mathbb{C}^n$, scalar $B>0, ~\bar{B}>0, ~\sigma>0$, The following inequality \eqref{xt} holds \cite{abughalwa2022finite}:
		\begin{align} \label{xt}
			\frac{||\mathbf{A}||^2}{B+\sigma}\geq \frac{||\bar{\mathbf{A}}||^2}{\bar{B}+\sigma}\left(2\frac{\Re\{\bar{\mathbf{A}}^H \mathbf{A}\}}{||\bar{\mathbf{A}}||^2}
			-\frac{B+\sigma}{\bar{B}+\sigma} \right).
		\end{align}
		
	\end{appendices}
	
	\bibliographystyle{IEEEtran}
	\bibliography{surface}

\begin{thebibliography}{10}
\providecommand{\url}[1]{#1}
\csname url@samestyle\endcsname
\providecommand{\newblock}{\relax}
\providecommand{\bibinfo}[2]{#2}
\providecommand{\BIBentrySTDinterwordspacing}{\spaceskip=0pt\relax}
\providecommand{\BIBentryALTinterwordstretchfactor}{4}
\providecommand{\BIBentryALTinterwordspacing}{\spaceskip=\fontdimen2\font plus
\BIBentryALTinterwordstretchfactor\fontdimen3\font minus \fontdimen4\font\relax}
\providecommand{\BIBforeignlanguage}[2]{{%
\expandafter\ifx\csname l@#1\endcsname\relax
\typeout{** WARNING: IEEEtran.bst: No hyphenation pattern has been}%
\typeout{** loaded for the language `#1'. Using the pattern for}%
\typeout{** the default language instead.}%
\else
\language=\csname l@#1\endcsname
\fi
#2}}
\providecommand{\BIBdecl}{\relax}
\BIBdecl

\bibitem{nguyen2022leveraging}
T.~V. Nguyen, D.~N. Nguyen, M.~D. Renzo, and R.~Zhang, ``Leveraging secondary reflections and mitigating interference in multi-{IRS}/{RIS} aided wireless networks,'' \emph{IEEE Transactions on Wireless Communications}, vol.~22, no.~1, pp. 502--517, 2023.

\bibitem{abughalwa2022finite}
M.~Abughalwa, H.~D. Tuan, D.~N. Nguyen, H.~V. Poor, and L.~Hanzo, ``Finite-blocklength {RIS}-aided transmit beamforming,'' \emph{IEEE Transactions on Vehicular Technology}, vol.~71, no.~11, pp. 12\,374--12\,379, 2022.

\bibitem{9896755}
G.~Chen, Q.~Wu, C.~He, W.~Chen, J.~Tang, and S.~Jin, ``Active {IRS} aided multiple access for energy-constrained {IoT} systems,'' \emph{IEEE Transactions on Wireless Communications}, vol.~22, no.~3, pp. 1677--1694, 2023.

\bibitem{zhou2021secure}
G.~Zhou, C.~Pan, H.~Ren, K.~Wang, and Z.~Peng, ``Secure wireless communication in {RIS}-aided {MISO} system with hardware impairments,'' \emph{IEEE Wireless Communications Letters}, vol.~10, no.~6, pp. 1309--1313, 2021.

\bibitem{10256584}
W.~Li, W.~Yu, H.~Liu, and H.~Hou, ``Robust secrecy rate maximization for {IRS}-aided {MISO} communication systems,'' in \emph{2023 IEEE 13th International Conference on CYBER Technology in Automation, Control, and Intelligent Systems (CYBER)}, 2023, pp. 604--609.

\bibitem{durisi2016toward}
G.~Durisi, T.~Koch, and P.~Popovski, ``Toward massive, ultrareliable, and low-latency wireless communication with short packets,'' \emph{Proceedings of the IEEE}, vol. 104, no.~9, pp. 1711--1726, 2016.

\bibitem{10904325}
W.~Zhao, J.~Ni, S.~Hao, B.~Li, and T.~Zhang, ``Joint physical layer security and information freshness of {RIS}-assisted {SPC} system: Analysis and optimization,'' \emph{IEEE Transactions on Vehicular Technology}, pp. 1--13, 2025.

\bibitem{10068292}
G.~Xie, C.~Yang, Y.~Feng, G.~Liu, and B.~Dai, ``Secure finite blocklength coding schemes for reconfigurable intelligent surface aided wireless channels with feedback,'' \emph{IEEE Transactions on Communications}, vol.~71, no.~5, pp. 2931--2946, 2023.

\bibitem{6802432}
W.~Yang, G.~Durisi, T.~Koch, and Y.~Polyanskiy, ``Quasi-static multiple-antenna fading channels at finite blocklength,'' \emph{IEEE Transactions on Information Theory}, vol.~60, no.~7, pp. 4232--4265, 2014.

\bibitem{10082954}
K.~Singh, S.~K. Singh, and C.-P. Li, ``On the performance analysis of {RIS}-assisted infinite and finite blocklength communication in presence of an eavesdropper,'' \emph{IEEE Open Journal of the Communications Society}, vol.~4, pp. 854--872, 2023.

\bibitem{9402750}
L.~Dong, H.-M. Wang, and H.~Xiao, ``Secure cognitive radio communication via intelligent reflecting surface,'' \emph{IEEE Transactions on Communications}, vol.~69, no.~7, pp. 4678--4690, 2021.

\bibitem{7529226}
G.~Durisi, T.~Koch, and P.~Popovski, ``Toward massive, ultrareliable, and low-latency wireless communication with short packets,'' \emph{Proceedings of the IEEE}, vol. 104, no.~9, pp. 1711--1726, 2016.

\bibitem{9197675}
F.~Fang, Y.~Xu, Q.-V. Pham, and Z.~Ding, ``Energy-efficient design of {IRS}-{NOMA} networks,'' \emph{IEEE Transactions on Vehicular Technology}, vol.~69, no.~11, pp. 14\,088--14\,092, 2020.

\bibitem{boyd1994linear}
S.~Boyd, L.~El~Ghaoui, E.~Feron, and V.~Balakrishnan, \emph{Linear matrix inequalities in system and control theory}.\hskip 1em plus 0.5em minus 0.4em\relax SIAM, 1994.

\bibitem{9525400}
H.~Niu, Z.~Chu, F.~Zhou, and Z.~Zhu, ``Simultaneous transmission and reflection reconfigurable intelligent surface assisted secrecy {MISO} networks,'' \emph{IEEE Communications Letters}, vol.~25, no.~11, pp. 3498--3502, 2021.

\bibitem{9963699}
A.~Al-Rimawi and A.~Al-Dweik, ``On the performance of {RIS}-assisted communications with direct link over $\kappa$-$\mu$ shadowed fading,'' \emph{IEEE Open Journal of the Communications Society}, vol.~3, pp. 2314--2328, 2022.

\bibitem{kammoun2020asymptotic}
Q.-U.-A. Nadeem, A.~Kammoun, A.~Chaaban, M.~Debbah, and M.-S. Alouini, ``Asymptotic {Max}-{Min} {SINR} analysis of reconfigurable intelligent surface assisted {MISO} systems,'' \emph{IEEE Transactions on Wireless Communications}, vol.~19, no.~12, pp. 7748--7764, 2020.

\bibitem{9110587}
G.~Zhou, C.~Pan, H.~Ren, K.~Wang, M.~D. Renzo, and A.~Nallanathan, ``Robust beamforming design for intelligent reflecting surface aided {MISO} communication systems,'' \emph{IEEE Wireless Communications Letters}, vol.~9, no.~10, pp. 1658--1662, 2020.

\bibitem{9501057}
B.~Zheng, C.~You, and R.~Zhang, ``Uplink channel estimation for double-irs assisted multi-user mimo,'' in \emph{ICC 2021 - IEEE International Conference on Communications}, 2021, pp. 1--6.

\bibitem{9373363}
------, ``Efficient channel estimation for double-irs aided multi-user mimo system,'' \emph{IEEE Transactions on Communications}, vol.~69, no.~6, pp. 3818--3832, 2021.

\bibitem{8579566}
C.~Pan, H.~Ren, M.~Elkashlan, A.~Nallanathan, and L.~Hanzo, ``Robust beamforming design for ultra-dense user-centric {C-RAN} in the face of realistic pilot contamination and limited feedback,'' \emph{IEEE Transactions on Wireless Communications}, vol.~18, no.~2, pp. 780--795, 2019.

\bibitem{9266086}
Z.~Zhang, L.~Lv, Q.~Wu, H.~Deng, and J.~Chen, ``Robust and secure communications in intelligent reflecting surface assisted {NOMA} networks,'' \emph{IEEE Communications Letters}, vol.~25, no.~3, pp. 739--743, 2021.

\bibitem{niu2021weighted}
H.~Niu, Z.~Chu, F.~Zhou, Z.~Zhu, M.~Zhang, and K.-K. Wong, ``Weighted sum secrecy rate maximization using intelligent reflecting surface,'' \emph{IEEE Transactions on Communications}, vol.~69, no.~9, pp. 6170--6184, 2021.

\bibitem{chu2020secrecy}
Z.~Chu, W.~Hao, P.~Xiao, D.~Mi, Z.~Liu, M.~Khalily, J.~R. Kelly, and A.~P. Feresidis, ``Secrecy rate optimization for intelligent reflecting surface assisted {MIMO} system,'' \emph{IEEE Transactions on Information Forensics and Security}, vol.~16, pp. 1655--1669, 2021.

\bibitem{9603291}
X.~Guan, Q.~Wu, and R.~Zhang, ``Anchor-assisted channel estimation for intelligent reflecting surface aided multiuser communication,'' \emph{IEEE Transactions on Wireless Communications}, vol.~21, no.~6, pp. 3764--3778, 2022.

\bibitem{8665906}
W.~Yang, R.~F. Schaefer, and H.~V. Poor, ``Wiretap channels: Nonasymptotic fundamental limits,'' \emph{IEEE Transactions on Information Theory}, vol.~65, no.~7, pp. 4069--4093, 2019.

\bibitem{feng2021reliable}
C.~Feng, H.-M. Wang, and H.~V. Poor, ``Reliable and secure short-packet communications,'' \emph{IEEE Transactions on Wireless Communications}, vol.~21, no.~3, pp. 1913--1926, 2021.

\bibitem{bloch2008wireless}
M.~Bloch, J.~Barros, M.~R.~D. Rodrigues, and S.~W. McLaughlin, ``Wireless information-theoretic security,'' \emph{IEEE Transactions on Information Theory}, vol.~54, no.~6, pp. 2515--2534, 2008.

\bibitem{TTN16}
H.~H.~M. Tam, H.~D. Tuan, and D.~T. Ngo, ``Successive convex quadratic programming for quality-of-service management in full-duplex {MU}-{MIMO} multicell networks,'' \emph{IEEE Transactions on Communications}, vol.~64, no.~6, pp. 2340--2353, 2016.

\bibitem{niu2022joint}
H.~Niu, Z.~Lin, Z.~Chu, Z.~Zhu, P.~Xiao, H.~X. Nguyen, I.~Lee, and N.~Al-Dhahir, ``Joint beamforming design for secure {RIS}-assisted {IoT} networks,'' \emph{IEEE Internet of Things Journal}, vol.~10, no.~2, pp. 1628--1641, 2023.

\bibitem{grant2014cvx}
M.~Grant and S.~Boyd, ``{CVX}: Matlab software for disciplined convex programming, version 2.1,'' 2014.

\bibitem{labit2002sedumi}
Y.~Labit, D.~Peaucelle, and D.~Henrion, ``{SeDuMi} interface 1.02: a tool for solving {LMI} problems with sedumi,'' in \emph{Proceedings. IEEE International Symposium on Computer Aided Control System Design}, 2002, pp. 272--277.

\bibitem{lipp2016variations}
T.~Lipp and S.~Boyd, ``Variations and extension of the convex--concave procedure,'' \emph{Optimization and Engineering}, vol.~17, pp. 263--287, 2016.

\bibitem{boyd2004convex}
S.~Boyd and L.~Vandenberghe, \emph{Convex optimization}.\hskip 1em plus 0.5em minus 0.4em\relax Cambridge university press, 2004.

\bibitem{1369660}
Y.~Eldar, A.~Ben-Tal, and A.~Nemirovski, ``Robust mean-squared error estimation in the presence of model uncertainties,'' \emph{IEEE Transactions on Signal Processing}, vol.~53, no.~1, pp. 168--181, 2005.

\bibitem{9774882}
W.~Wang, W.~Ni, H.~Tian, Z.~Yang, C.~Huang, and K.-K. Wong, ``Safeguarding {NOMA} networks via reconfigurable dual-functional surface under imperfect {CSI},'' \emph{IEEE Journal of Selected Topics in Signal Processing}, vol.~16, no.~5, pp. 950--966, 2022.

\bibitem{9505311}
Y.~Chen, Y.~Wang, and L.~Jiao, ``Robust transmission for reconfigurable intelligent surface aided millimeter wave vehicular communications with statistical {CSI},'' \emph{IEEE Transactions on Wireless Communications}, vol.~21, no.~2, pp. 928--944, 2022.

\bibitem{9180053}
G.~Zhou, C.~Pan, H.~Ren, K.~Wang, and A.~Nallanathan, ``A framework of robust transmission design for {IRS}-aided {MISO} communications with imperfect cascaded channels,'' \emph{IEEE Transactions on Signal Processing}, vol.~68, pp. 5092--5106, 2020.

\bibitem{6772207}
A.~D. Wyner, ``The wire-tap channel,'' \emph{The Bell System Technical Journal}, vol.~54, no.~8, pp. 1355--1387, 1975.

\bibitem{BOL19}
E.~Bj{\H o}rnson, {\H O}.~{\H O}zdogan, and E.~G. Larsson, ``Intelligent reflecting surface versus decode-and-forward: How large surfaces are needed to beat relaying?'' \emph{IEEE Wireless Communications Letters}, vol.~9, no.~2, pp. 244--248, 2020.

\bibitem{ben18}
M.~Bennis, M.~Debbah, and H.~V. Poor, ``Ultrareliable and low-latency wireless communication: Tail, risk, and scale,'' \emph{Proc. IEEE}, vol. 106, no.~10, pp. 1834--1853, 2018.

\bibitem{sh17cross}
C.~She, C.~Yang, and T.~Q. Quek, ``Cross-layer optimization for ultra-reliable and low-latency radio access networks,'' \emph{IEEE Trans. Wirel. Commun.}, vol.~17, no.~1, pp. 127--141, 2017.

\bibitem{jain1984quantitative}
R.~K. Jain, D.-M.~W. Chiu, W.~R. Hawe \emph{et~al.}, ``A quantitative measure of fairness and discrimination,'' \emph{Eastern Research Laboratory, Digital Equipment Corporation, Hudson, MA}, vol.~21, 1984.

\bibitem{vmimo}
\BIBentryALTinterwordspacing
D.~N. Nguyen and M.~Krunz, ``A cooperative mimo framework for wireless sensor networks,'' \emph{ACM Trans. Sen. Netw.}, vol.~10, no.~3, May 2014. [Online]. Available: \url{https://doi.org/10.1145/2499381}
\BIBentrySTDinterwordspacing

\end{thebibliography}
\end{document}